\newif\ifsingle

\newif\ifproofs

\newif\ifFullVersion

\ifsingle
\documentclass[11pt,draftclsnofoot, onecolumn]{IEEEtran}		
\else		
\documentclass[10pt,final, twocolumn]{IEEEtran}
\fi

\IEEEoverridecommandlockouts
\usepackage{balance}
\usepackage{amsmath}
\usepackage{amsthm}
\usepackage[bookmarks,colorlinks]{hyperref}
\usepackage{cite}
\usepackage{cases}
\usepackage{amssymb}
\usepackage{algorithm}
\usepackage{algorithmic}
\usepackage{graphicx}
\usepackage{caption}
\usepackage{subcaption}
\usepackage{color}
\usepackage{enumerate}
\usepackage{xcolor}
\usepackage[bookmarks,colorlinks]{hyperref}
\usepackage{acronym}
\usepackage{epstopdf}

\newtheorem{thm}{Theorem}
\newtheorem{lem}{Lemma}
 
\newtheorem{corollary}{Corollary}
\newtheorem{proposition}{Proposition}

\definecolor{NewColor}{rgb}{0, 0, 0}

\acrodef{adc}[ADC]{analog-to-digital convertor}
\acrodef{dft}[DFT]{discrete Fourier transform}
\acrodef{cs}[CS]{compressed sensing}
\acrodef{bs}[BS]{base station}
\acrodef{rf}[RF]{radio frequency}
\acrodef{ula}[ULA]{uniform linear array}
\acrodef{em}[EM]{electromagnetic}
\acrodef{los}[LOS]{line-of-sight}
\acrodef{nlos}[NLOS]{non-line-of-sight}
\acrodef{lfm}[LFM]{linear frequency modulated}
 
\ifsingle

\else

\fi 

\DeclareMathOperator{\diag}{diag}

\DeclareMathOperator{\sign}{sign}
\DeclareMathOperator{\Tr}{Tr}
\DeclareMathOperator{\rank}{rank}

\IEEEoverridecommandlockouts

\begin{document}
%
\title{Near-Field Channel Estimation and Joint Angle-Range Recovery in XL-MIMO Systems: A Gridless Super-Resolution Approach}
%
%
%

\author{Feng~Xi,~\IEEEmembership{Member,~IEEE,}
        Dehui~Yang
\thanks{F. Xi is with the Department
of Electronic Engineering, Nanjing University of Science and Technology, Nanjing 210094, China (email: xifeng@njust.edu.cn). 
D. Yang is with the School of Mathematics and Physics, Xi’an Jiaotong-Liverpool University (XJTLU), Suzhou 215123, China (email: dehui.yang@xjtlu.edu.cn).
The work of F. Xi  was supported in part by the National Science Foundation of China  under grant No. 62471230. D. Yang was supported in part by the
Research Development Fund (RDF-23-02-080) from XJTLU. ({\em Corresponding author: Dehui Yang})}
}

\maketitle
\pagestyle{plain}
\thispagestyle{plain}

\begin{abstract}
Existing near‑field channel estimation methods for extremely large‑scale MIMO (XL‑MIMO) typically discretize angle and range parameters jointly, resulting in large polar‑domain codebooks. This paper proposes a novel framework that formulates near‑field channel estimation as a gridless super‑resolution problem, eliminating the need for explicitly constructed codebooks. By employing a second‑order approximation of spherical‑wave steering vectors, the near-field channel is represented as a superposition of complex exponentials modulated by unknown waveforms. We demonstrate that these waveforms lie tightly in a common discrete chirp rate (DCR) subspace, with a dimension that scales as $\Theta(\sqrt{N})$ for an $N$-element array. 
By leveraging this structure and applying a lifting technique, we reformulate the non-convex problem as a convex program using regularized atomic norm minimization, which admits an equivalent semidefinite program. 
From the solution to the convex program, we obtain gridless angle estimates and derive closed‑form coarse range estimates, followed by refinement under the exact spherical model using gradient‑based nonlinear least squares. The proposed method avoids basis mismatch and exhaustive two‑dimensional grid searches while enabling accurate joint angle-range estimation with pilot budgets that scale sublinearly with array size in sparse multipath regimes. Simulations demonstrate accurate channel reconstruction and user localization across representative near‑field scenarios.
\end{abstract}

\begin{IEEEkeywords}
Near‑field channel estimation, super‑resolution, atomic norm minimization, joint angle-range estimation
\end{IEEEkeywords}

%
\IEEEpeerreviewmaketitle

\section{Introduction} 
\label{sec:intro}
\vspace{-0.1cm}
\IEEEPARstart{W}{ith} the rapid evolution of sixth-generation (6G) wireless communications \cite{Saad-MNET2020}, innovative paradigms such as extremely large-scale multiple-input multiple-output (XL-MIMO) \cite{Wang-COMST2024}, holographic MIMO \cite{Gong-COMST2024}, and reconfigurable intelligent surfaces (RISs) \cite{Pan-MCOM2021} have emerged as key enabling technologies for enhancing network capacity, expanding coverage, and improving reliability.
Moreover, as integrated sensing and communication (ISAC) \cite{Liu-2022JSAC} plays a pivotal role in 6G architectures, it is imperative to develop communication systems that deliver high-resolution sensing and accurate localization capabilities.
To meet these requirements, 6G research is expanding beyond the current mid-band spectrum to  millimeter-wave and sub-terahertz  frequencies, achieving significantly higher spectral efficiency.
However, these technologies and stringent performance requirements introduce significant challenges for channel estimation—a fundamental task in wireless communications.

Channel estimation aims to acquire accurate channel state information (CSI), which is essential for various optimization tasks, such as maximizing spectral efficiency and enabling precise user localization.
Traditionally, channels are modeled under the far-field assumption, and a diverse range of estimation methods, spanning both parametric and non-parametric approaches, have been developed \cite{Wan-TSP2008,Rodriguez-2018TWC,Lee-TCOMM2016,Huang-TSP2019}.
However, as the number of antenna elements increases to massive scales and the utilized spectrum shifts to higher frequencies, the far-field model, which assumes planar-wave propagation, becomes inadequate for characterizing the communication channel.
Consequently, a paradigm shift from the far-field planar-wave assumption to the near-field spherical-wave model is necessary. However, this introduces substantial challenges in acquiring accurate near-field CSI.

The boundary between the near-field and far-field regions is determined by the Rayleigh distance \cite{Selvan-2017APM}, which grows quadratically with the array aperture and decreases inversely with the wavelength.
Thus, in 6G systems employing XL-MIMO arrays, users are highly likely to reside within the near-field region.
In this regime, the channel exhibits a high-dimensional structure due to the large number of antennas, rendering classical least-squares estimation prohibitively demanding in terms of pilot overhead.
Therefore, exploiting the intrinsic structure of near-field channels is crucial for efficient CSI acquisition.
However, the extensively leveraged angular-domain sparsity for far-field channel estimation \cite{Lee-TCOMM2016,Rodriguez-2018TWC} is unsuitable for near-field channels, due to the severe energy spread effect induced by spherical wavefronts \cite{Cui-ToC2022}. 
One solution for addressing this issue is to partition the XL array into several sub-arrays such that each sub-array operates in a local far-field, allowing existing angular-domain methods to remain applicable \cite{Amiri-GC2018,Han-WCL2020,Pisharody-TWC2024}.
Nevertheless, this approach requires inter-sub-array coordination, significantly complicating the channel estimation task.

A more principled approach is the use of a polar-domain sparse representation \cite{Cui-ToC2022}, which jointly parameterizes each propagation path by its angle and range parameters.
By constructing a codebook over discretized angle-range grids and employing compressive sensing and sparse recovery techniques, the CSI of the near-field channel can then be accurately recovered using significantly shorter pilot sequences.
Building upon this framework, various near-field channel estimation methods have been developed for diverse 6G applications, including wideband near-field channels \cite{Cui-SCIS2023,Ruan-TVT2025}, non-stationary channels \cite{Chen-TWC2024,Zhang-WCL2024}, and hybrid-field channels \cite{Wei-LCOMM2022,Lei-TSP2024}, among others.
These studies demonstrate the efficiency of the polar-domain codebook for near-field channel representation.
However, discretizing the angle-range space introduces columns with high correlation and substantial computational overhead \cite{Abdallah-SPM2025}.
To mitigate these issues, several alternative codebooks have recently been proposed, such as the spatial-chirp codebook \cite{Shi-TWC2024}, 
the discrete prolate spheroidal sequence (DPSS)-based codebook \cite{Liu-ICC2024}, 
and dictionary-learning designs that iteratively update the codebook while reconstructing the channel matrix \cite{Zhang-2024ToC}. 
Nevertheless, all existing designs rely on parameter discretization, and therefore remain vulnerable to basis mismatch arising from the inconsistencies between the discrete codebook and the underlying continuous-valued channel parameters.
Although off-grid refinement methods \cite{Cui-ToC2022,Lei-TSP2024} partially mitigate this issue, their performance is sensitive to the accuracy of the initial on-grid estimates.
Recently, there also exist works that employ atomic norm minimization (ANM) to avoid basis mismatch \cite{Rinchi-Globecom22,Daei-ICASSP25}. However, these methods neither directly recover the underlying near-field channel nor provide theoretical recovery guarantees. Instead, they extract only the angular parameters from the ANM solution and rely on iterative procedures or exhaustive searches to estimate the range parameters.

In addition to sparse recovery methods, several studies have adopted MUSIC-based subspace techniques \cite{Kosasih-Asilomar2023} and maximum likelihood estimation (MLE) \cite{Haghshenas-TWC2024} to directly retrieve the angle and range parameters of near-field propagation paths. Although these approaches circumvent explicit codebook construction, they typically require lengthy pilot sequences to form reliable sample covariance matrices or likelihood functions. Moreover, both subspace and MLE approaches still rely on exhaustive grid searches to locate spectrum peaks or maximize the likelihood surface, imposing significant computational overhead.
More recently, data-driven methods \cite{Jang-TCOMM2024,Ye-TGCN2025,Yuan-TWC2025} have been introduced to address the complexity of spherical wavefronts.
While these models excel at capturing intricate channel characteristics, they demand substantial computational resources and tend to generalize poorly in highly dynamic environments.
In summary, an efficient near-field channel estimation framework that avoids discretized codebooks while simultaneously delivering accurate CSI recovery and precise angle-range estimation for each propagation path remains elusive.

\vspace{-0.38cm}
\subsection{Main Contributions}
\label{subsec:Contribution}
This paper proposes a novel super-resolution framework for near-field channel estimation in XL-MIMO systems.
Our approach replaces the discretized polar-domain representation with a gridless atomic representation. 
The main contributions of this work are summarized as follows:
\begin{itemize}
    \item By employing a second-order Taylor expansion of the near-field array response, we demonstrate that the near-field channel vector consists of a set of complex exponentials modulated by unknown waveforms.
    Consequently, the near-field channel estimation is equivalent to resolving the set of complex exponentials and their modulated unknown waveforms, yielding a super-resolution problem.
    \item To address the resulting super-resolution problem for near-field channel estimation, we analyze the characteristics of the unknown modulating waveforms. Our analysis reveals that these waveforms lie within a common, known low-dimensional subspace, which can be effectively constructed from the \emph{discrete chirp rate (DCR)} dictionary. In particular, we show that a $\Theta(\sqrt{N})$-dimensional DCR subspace is sufficient to accurately represent the length-$N$ unknown modulated waveforms.
    \item By leveraging the sparsity of the near-field channel and the low-dimensionality of the DCR subspace representation, we reformulate the near-field channel estimation problem as a convex program using regularized atomic norm minimization, which can be efficiently solved using semidefinite programming (SDP). Furthermore, we provide a theoretical analysis of the recovery error in the presence of measurement noise for the proposed atomic norm minimization. The analysis demonstrates that our approach can robustly recover the underlying structured low-rank matrix, and thus the near-field channel and its associated parameters.
    \item To further enhance the accuracy of the proposed algorithm, we incorporate a gradient descent-based refinement step to iteratively optimize the angle and range parameters, mitigating residual errors due to the Taylor expansion of the near-field array response. Extensive simulations are conducted to verify the effectiveness of the proposed approaches and confirm the theoretical results. Simulation results demonstrate that our method not only provides accurate channel estimation but also effectively recovers both the angle and range parameters in the near-field region.
\end{itemize}

\vspace{-0.35cm}
\subsection{Notation and Organization of the Paper}
\label{subsec:Organization}
Throughout the paper,  
we use lowercase (uppercase) bold characters to denote vectors (matrices).
The $i$th element of a vector $\mathbf{x}$ is written as $[\mathbf{x}]_i$.
Similarly, the $(i,j)$th element of a matrix $\mathbf{X}$ is denoted as $[\mathbf{X}]_{i,j}$.
For a vector $\mathbf{x}$, $\mathbf{x}^T$, $\overline{\mathbf{x}}$, $\mathbf{x}^H$, and $\|\mathbf{x}\|_2$ denote its transpose, complex conjugate, conjugate transpose, and $\ell_2$ norm, respectively.
For a matrix $\mathbf{X}$, $\mathbf{X}^T$, $\overline{\mathbf{X}}$, $\mathbf{X}^H$, $\|\mathbf{X}\|_2$, $\|\mathbf{X}\|_F$, $\rank(\mathbf{X})$, and $\Tr(\mathbf{X})$ denote the transpose, complex conjugate, conjugate transpose, operator norm, Frobenius norm, rank, and trace of $\mathbf{X}$, respectively.
The inner product of matrices $\mathbf{X}$ and $\mathbf{Y}$ is defined as $\langle\mathbf{X},\mathbf{Y}\rangle = \text{Tr}(\mathbf{Y}^H\mathbf{X})$ and $\langle \mathbf{X}, \mathbf{Y}\rangle_{\mathbb{R}} = \text{Re}(\langle \mathbf{X}, \mathbf{Y}\rangle)$.
$\mathbf{X}\succeq 0$ indicates that $\mathbf{X}$ is positive
semidefinite (PSD).
We use $[N]$ and $\mathbf{I}_N$ to denote the set $\{0,1,\cdots,N-1\}$ and an $N \times N$ identity matrix, respectively.
$\mathbb{R}$ and $\mathbb{C}$ denote the sets of real and complex numbers.

The rest of the paper is organized as follows. Section \ref{sec:Model} introduces the near-field channel model and characteristics of the corresponding steering vectors. In Section \ref{sec:method}, we formulate the channel estimation problem as a gridless super-resolution and parameter estimation task using a lifting technique and atomic norm minimization. In Section \ref{sec:ParameterEstimation}, we propose a two-stage scheme for angle and range parameter estimation and refinement.   Section \ref{sec:Sims} provides numerical simulations to support and illustrate the superior performance of our approach. Finally,
concluding remarks are given in Section \ref{sec:conclusion}.

\begin{figure}[htbp]
\centerline{\includegraphics[width=0.48\textwidth]{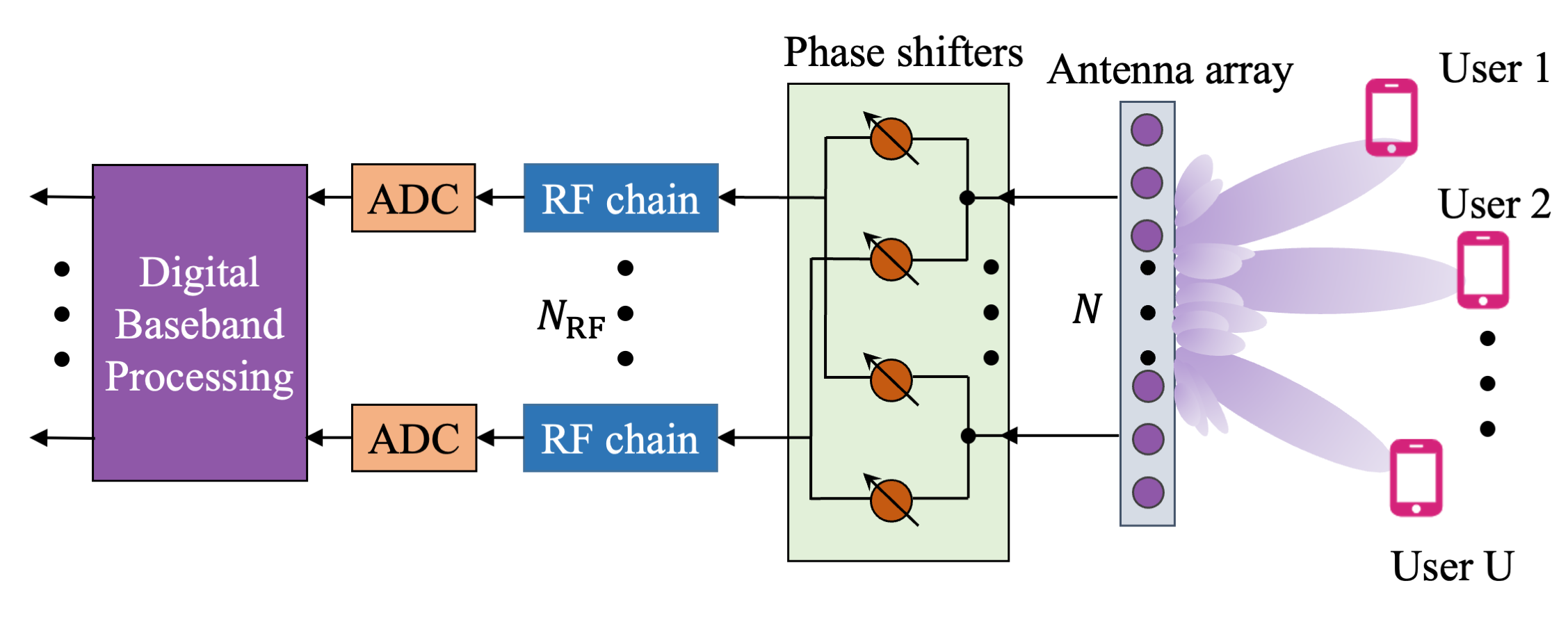}}
\caption{An XL-MIMO system with a hybrid analog and digital transceiver architecture.}
\label{fig:MIMOSystem}
\end{figure}

\section{System and Channel Model}
\label{sec:Model}
\vspace{-0.1cm}
This section introduces the XL-MIMO system model and  provides a detailed characterization of the near-field channel structure.

\subsection{System Model}
\label{subsec:Model}
Consider an uplink multi-user XL-MIMO communication system serving $U$ single-antenna users, as shown in Fig. \ref{fig:MIMOSystem}. The \ac{bs} employs a hybrid analog and digital transceiver architecture to reduce the number of required \ac{rf} chains.
Specifically, the BS is equipped with a \ac{ula} comprising $N$ antennas and $N_{\text{RF}}$ \ac{rf} chains, where $N_{\text{RF}}\ll N$. Each \ac{rf} chain is connected to the antenna array via analog phase shifters. The antenna elements are spaced by $d=\lambda_c/2$, where $\lambda_c$ denotes the carrier wavelength.
Geometrically, 
the antennas are placed along the $y$-axis, with the $n$th antenna located at position $(0,(n-1)d)$, $n=1,\cdots,N$. 
During uplink channel estimation, $U$ users transmit mutually orthogonal pilot sequences to the \ac{bs}, with $U\leq N_{\text{RF}}$, enabling independent channel estimation for each user \cite{Emil-2016CM}. 

Without loss of generality, consider an arbitrary user located at $(x,y)$ with polar coordinates $(r,\theta) = (\sqrt{x^2+y^2}, \arctan \frac{y}{x})$. Each user transmits a pilot sequence $\{s_1, s_2,\cdots, s_Q\}$ of length $Q$ during the channel estimation phase, where $s_q$ denotes the pilot signal transmitted in the $q$th time slot. 
The received pilot signal during the $q$th time slot at the \ac{bs} is given by
\begin{equation}
    \widetilde{\mathbf{y}}_q = \mathbf{A}_q(\mathbf{h}s_q + \mathbf{n}_q)\in\mathbb{C}^{N_{\text{RF}}\times 1},
\end{equation}
where $\mathbf{A}_q\in\mathbb{C}^{N_{\text{RF}}\times N}$ denotes the analog combining matrix, $\mathbf{h}\in\mathbb{C}^{N\times 1}$ represents the channel vector between the \ac{bs} and each user, and $\mathbf{n}_q\in\mathbb{C}^{N\times 1}$ denotes the additive noise during signal transmission and acquisition. We assume $\mathbf{n}_q\sim\mathcal{CN}(0,\sigma^2 \mathbf{I}_N)$, where $\sigma^2$ is the noise variance.
The analog combining matrix $\mathbf{A}_q$ is typically implemented via analog phase shifters and satisfies the constant modulus constraint, i.e., $|[\mathbf{A}_q]_{i,j}| = \frac{1}{\sqrt{N}}$.

Denoting $\mathbf{y}_q = \frac{\overline{s_q}}{|s_q|^2}\widetilde{\mathbf{y}}_q$ and $\mathbf{y}=[\mathbf{y}_1^{T},\cdots,\mathbf{y}_Q^{T}]^{T}\in\mathbb{C}^{N_{\text{RF}}Q\times 1}$, we obtain the following stacked observation model
\begin{equation}\label{eqn:y}
    \mathbf{y} = \mathbf{A}\mathbf{h} + \mathbf{n},
\end{equation}
where $\mathbf{A} = [\mathbf{A}_1^{T},\cdots,\mathbf{A}_Q^{T}]^{T}\in\mathbb{C}^{N_{\text{RF}}Q\times N}$ and $\mathbf{n}=[(\frac{\overline{s_1}}{|s_1|^2}\mathbf{A}_1\mathbf{n}_1)^{T},\cdots,(\frac{\overline{s_Q}}{|s_Q|^2}\mathbf{A}_Q\mathbf{n}_Q)^{T}]^{T}\in\mathbb{C}^{N_{\text{RF}}Q\times 1}$.

In large-scale MIMO systems, the number of antennas $N$ at the \ac{bs} is typically much larger than the number of available measurements $N_{\text{RF}}Q$, making the channel estimation problem severely underdetermined.
To address this ill-posed problem for channel estimation, it is crucial to exploit the intrinsic structural properties of the channel vector $\mathbf{h}$.


\subsection{Near-Field Channel Model}
\label{subsec:Channel}
To characterize the near-field channel, we first recall two well-established distance boundaries that partition the \ac{em} radiation field into distinct regions.

The first is the \textit{Rayleigh distance}, also known as the \textit{Fraunhofer distance}, which separates the far-field and near-field regions \cite{Selvan-2017APM}. 
It arises from the curvature of the wavefront.
Specifically, in the far-field regime, the phase of an \ac{em} wave can be accurately approximated by a first-order Taylor expansion, resulting in negligible phase error.
As the distance decreases, the residual phase error increases.
The Rayleigh distance is defined as the location where the phase error reaches $\pi/8$, given by $Z_{ra}=\frac{2D^2}{\lambda_c}$, where $D$ denotes the array aperture.
For a communication distance $r > Z_{ra}$, the planar-wave model is valid; otherwise, spherical-wave effects dominate.

The second boundary is the \textit{Fresnel distance}, given by $Z_{fr}=0.62\sqrt{\frac{D^3}{\lambda_c}}$, which further partitions the near-field region into the \textit{radiating near-field region} (also known as the \textit{Fresnel region}) and the \textit{reactive near-field region}.
In the radiating near-field region, i.e., $Z_{fr} <r<Z_{ra}$, a second-order Taylor expansion accurately captures the wavefront curvature.
In the reactive near-field region, i.e., $r<Z_{fr}$, the wavefront is dominated by evanescent components, and the field structure becomes significantly more complex.
Since $Z_{fr}$ is typically much smaller than $Z_{ra}$, most practical XL-MIMO communication scenarios only involve users in the radiating near-field region. Throughout the remainder of this paper, we restrict our analysis to this regime.



Under the spherical wavefront assumption, the near-field channel vector is modeled as
\begin{equation}\label{eqn:h}
    \mathbf{h}=\sum_{l=1}^L \widetilde{\alpha}_l e^{-j2\pi\frac{r_l}{\lambda_c}} \mathbf{b}(\theta_l, r_l),
\end{equation}
where $L$ is the total number of propagation paths, $\widetilde{\alpha}_l$ is the complex gain of the $l$th path, and $\theta_l$ and $r_l$ represent the angle and range of the $l$th path relative to the reference antenna. 
The near-field steering vector $\mathbf{b}(\theta_l, r_l)$ is defined as
\begin{equation}\label{eqn:b_theta_r}
    \mathbf{b}(\theta_l, r_l) = \left[e^{-j2\pi \frac{(r^{(1)}_l-r_l)}{\lambda_c}}, \cdots,e^{-j2\pi \frac{(r^{(N)}_l-r_l)}{\lambda_c}} \right]^T,
\end{equation}
where $r^{(n)}_l$ denotes the distance between the $n$th BS antenna and the user or scatterer.
Since the $n$th BS antenna is located at $(0,(n-1)d)$, 
$r^{(n)}_l$ is given by
\begin{equation}\label{eqn:rn_approx}
\begin{split}
     r^{(n)}_l &= \sqrt{r_l^2 + ((n-1)d)^2 - 2r_l(n-1)d\sin{\theta_l}}\\
     &\overset{(a)}{\approx} r_l -(n-1)d\sin{\theta_l} + \frac{(n-1)^2d^2\cos^2{\theta_l}}{2r_l},
\end{split}
\end{equation}
where the approximation $(a)$ follows from the second-order Taylor expansion $\sqrt{(1+x)}\approx 1+ \frac{1}{2}x - \frac{1}{8}x^2 + \mathcal{O}(x^3)$.
As $r_l\to\infty$, the steering vector degenerates to the far-field form.
Without loss of generality, we assume that the first path corresponds to the \ac{los} path, while the remaining ones are \ac{nlos} paths. 

The near-field steering vector depends on both the angle and range parameters, leading to a more intricate structure that makes the channel estimation problem challenging.
To leverage this structure, \cite{Cui-ToC2022} introduced the polar-domain representation, where the near-field channel is approximated as
\begin{equation}
    \mathbf{h}=\mathbf{D}\mathbf{u},
\end{equation}
where $\mathbf{u}$ is the sparse representation vector, and $\mathbf{D}$ is an angle-range dictionary constructed over a set of discretized grids $\{(\theta_{1},r_{1}^1),\cdots, (\theta_{1},r_{1}^{S_1}),\cdots, (\theta_{K},r_{K}^1), \cdots,  (\theta_{K},r_{K}^{S_K})\}$ with $S_k$ representing the number of sampled ranges for the sampled angle $\theta_k$.
Thus, the total number of sampled grids is $\sum_{k=1}^K S_k$.
While this method effectively approximates the near-field channel vector, it suffers from two drawbacks: (i) a prohibitively large dictionary size, which imposes significant storage and computational burdens, and (ii) inherent off-grid issues due to discretization, which degrade estimation and localization accuracy.
In this work, we circumvent these limitations by directly exploiting the intrinsic parametric structure of the near-field channel, recovering $\mathbf{h}$ and the associated angle-range parameters $\{(\theta_l, r_l)\}_{l=1}^L$ without constructing a two-dimensional discretized dictionary $\mathbf{D}$.

\section{A Super-Resolution Framework for Near-Field Channel Estimation}
\label{sec:method}
This section develops a super-resolution framework for near-field channel estimation. We first reformulate the near-field channel as a superposition of complex exponentials modulated by unknown waveforms. Furthermore, we demonstrate that these unknown waveforms lie approximately in a low-dimensional discrete chirp rate subspace, and integrate this subspace structure into a regularized atomic norm minimization formulation for near-field channel estimation. Finally, we provide theoretical analysis and guarantees of the proposed super-resolution approach for near-field channel recovery.

\vspace{-0.2cm}
\subsection{Gridless Super-Resolution Formulation}
\label{subsec:BSR}
\vspace{-0.1cm}

Substituting the second-order approximation  (\ref{eqn:rn_approx}) into (\ref{eqn:b_theta_r}), the near-field steering vector $\mathbf{b}(\theta_l,r_l)$ can be approximated as
\begin{equation}\label{eqn: b_theta_r_appro}
    \mathbf{b}(\theta_l,r_l)\approx\left[1, e^{j2\pi[\omega_l - \phi_l]}, \cdots,e^{j2\pi[(N-1)\omega_l - (N-1)^2\phi_l]} \right]^T,
\end{equation}
where $\omega_l = \frac{d \sin{\theta_l}}{\lambda_c}$ and $\phi_l = \frac{ d^2\cos^2{\theta_l}}{2\lambda_c r_l}$.

For notational simplicity, define two vectors:
\begin{equation}
\label{eqn:g}
\begin{cases}
\mathbf{a}(\omega) = [1,e^{j2\pi\omega},\cdots,e^{j2\pi(N-1)\omega}]^T\in \mathbb{C}^N,\\
\mathbf{g}(\phi) = [1,e^{j2\pi\phi},\cdots,e^{j2\pi(N-1)^2\phi}]^H \in \mathbb{C}^N.
\end{cases}
\end{equation}
Thus, the near-field steering vector $\mathbf{b}(\theta,r)$ can be written as
\begin{equation}
    \mathbf{b}(\theta,r) \approx \mathbf{a}(\omega)\odot \mathbf{g}(\phi),
\end{equation}
where $\odot$ denotes the Hadamard product, $\mathbf{a}(\omega)$ is the far-field steering vector depending only on the angle $\theta$, and $\mathbf{g}(\phi)$ is an unknown waveform determined jointly by the angle and range parameters $(\theta, r)$. Thus, the near-field channel vector $\mathbf{h}$ from (\ref{eqn:h}) can be rewritten as
\begin{equation}\label{eqn:h_BSR}
    \mathbf{h} \approx \sum_{l=1}^L \alpha_l (\mathbf{a}(\omega_l) \odot \mathbf{g}(\phi_l)),
\end{equation}
where we have defined $\alpha_l = \widetilde{\alpha}_l e^{-j2\pi\frac{r_l}{\lambda_c}}$.

From (\ref{eqn:h_BSR}), we observe that the near-field channel $\mathbf{h}$ is decomposed as a superposition of complex exponentials, parameterized by $\{\alpha_l,\omega_l\}_{l=1}^L$ and modulated by unknown waveforms $\{\mathbf{g}(\phi_l)\}_{l=1}^L$. 
In \cite{Yang-2016TIT}, to solve the inverse problem, the authors assume the unknown waveforms lie in a common and known low-dimensional subspace.
In our setting, the waveform $\mathbf{g}(\phi_l)$ is a sampled chirp signal with zero initial frequency and a frequency rate of $\phi_l$.

To construct a subspace for $\mathbf{g}(\phi_l)$, one straightforward approach is to discretize the chirp rate parameter $\phi$ and form a dictionary matrix based on the grid $\{\widetilde{\phi}_i\}_{i=1}^{N_\phi}$, where $N_\phi$ denotes the total number of grid points. When each $\mathbf{g}(\phi_l)$ is well-approximated by a small number of dictionary elements, we obtain a compact representation for $\mathbf{g}(\phi_l)$. Note that the feasibility of this discretization-based approach was demonstrated in our conference paper \cite{Li-ICASSP2025}. However, our prior work does not provide a theoretical guarantee on the approximation accuracy. Furthermore, the required subspace dimensionality to ensure accurate representation of the unknown waveforms remains unclear. To address these challenges, in this paper we introduce the discrete chirp rate subspace for the set of modulating waveforms $\{\mathbf{g}(\phi_l)\}_{l=1}^L$ and provide theoretical results characterizing both the dimensionality of the subspace and the approximation error when representing $\mathbf{g}(\phi)$.

\vspace{-0.2cm}
\subsection{Discrete Chirp Rate Subspace}
\label{subsec:DCR}
\vspace{-0.1cm}
To construct a low-dimensional subspace that captures the unknown waveforms $\{\mathbf{g}(\phi_l)\}_{l=1}^L$, we examine the structural properties of $\mathbf{g}(\phi)$ defined in (\ref{eqn:g}).
From the definition of $\phi$ in (\ref{eqn: b_theta_r_appro}), we observe that $\phi$ is jointly determined by the angle $\theta$ and the range $r$. Since the user distance is restricted to the radiating near-field region, i.e., $Z_{fr}<r<Z_{ra}$, the feasible range of $\phi$ can be bounded, as shown in the following lemma.

\begin{lem}\label{lemma1}
    For an arbitrary user located in the radiating near-field region of an $N$-element \ac{ula} with the antenna elements spaced by $d$, the parameters $\{\phi_l\}_{l=1}^L$ in the near-field channel model (\ref{eqn:h_BSR}) satisfy $\phi_l\in[0,\phi_{\max}]$, $l=1,\cdots, L$, where $\phi_{\max} = \frac{c}{(N-1)^{3/2}}$ 
    and $c = \frac{1}{1.24}\sqrt{\frac{d}{\lambda_c}}$ is a constant.
\end{lem} 

\begin{IEEEproof}
Since $\phi = \tfrac{d^2 \cos^2\theta}{2\lambda_c r}$, 
we have $\phi_{\max} = \frac{d^2}{2\lambda_c r_{\min}}$, where $r_{\min}$ is the minimum user distance. As users are restricted to the radiating near-field region, we have $r_{\min}=Z_{fr}=0.62\sqrt{D^3/\lambda_c}$, where the array aperture $D=(N-1)d$.
Substituting $r_{\min}$ into $\phi_{\max} = \frac{d^2}{2\lambda_c r_{\min}}$ completes the proof.
\end{IEEEproof}

Lemma~\ref{lemma1} shows that the admissible chirp rate interval shrinks polynomially with the array size $N$.
Thus, the set of waveforms $\{\mathbf{g}(\phi_l)\}_{l=1}^L$ is confined to a compact chirp rate interval, facilitating low-dimensional representation.
Specifically, this bounded region can be exploited to construct a set of discretized chirp rate grids and, subsequently, form a low-dimensional subspace where the set of unknown waveforms $\{\mathbf{g}(\phi_l)\}_{l=1}^L$ can be well-approximated.

Divide the interval $[0,\phi_{\max}]$ uniformly into $P$ ($P\geq2$) grid points $\widetilde{\phi}_i =(i-1)\phi_{\max}/(P-1)$, $i=1,\cdots,P$, 
and construct the normalized waveform $\mathbf{u}(\phi) = \mathbf{g}(\phi)/\sqrt{N}$. We then construct the \emph{DCR subspace} as 
\begin{equation}\label{eqn:DCR_subspace}
    \mathcal{V} = \mathrm{span} \{\mathbf{u}(\widetilde{\phi}_1),\cdots,\mathbf{u}(\widetilde{\phi}_P)\}.
\end{equation}
The approximation error of representing any waveform $\mathbf{g}(\phi), \phi\in[0,\phi_{\max}]$ using the subspace $\mathcal{V}$ is established in
the following proposition.

\begin{proposition}\label{Prop1}
    For any $\phi\in[0,\phi_{\max}]$, the relative approximation error of representing $\mathbf{g}(\phi)$ by the DCR subspace $\mathcal V$ satisfies
    \begin{equation}\label{eqn:dist_bound_exact}
    \frac{\mathrm{dist}\big(\mathbf{g}(\phi),\mathcal V\big)}{\|\mathbf{g}(\phi)\|_2}
    = \mathrm{dist}\big(\mathbf{u}(\phi),\mathcal V\big)
    \;\le\; \sqrt{\frac{\pi\,\delta\,(N-1)(2N-1)}{3}},
    \end{equation}
    where $\mathrm{dist}(\mathbf{x},\mathcal V) =  \inf_{\mathbf{v}\in\mathcal V}\|\mathbf{x}-\mathbf{v}\|_2$ and $\delta=\phi_{\max}/(P-1)$.
    A sufficient condition for $\mathrm{dist}\big(\mathbf{g}(\phi),\mathcal V\big)/\|\mathbf{g}(\phi)\|_2\le \epsilon$ uniformly for any $\phi\in[0,\phi_{\max}]$ is
    \begin{equation}\label{eqn:P_bound_exact}
    P \;\ge\; 1 \;+\; \frac{\pi\,\phi_{\max}\,(N-1)(2N-1)}{3\,\epsilon^2}\,.
    \end{equation}
    Since $\phi_{\max} = \frac{c}{(N-1)^{3/2}}$ as shown in Lemma \ref{lemma1}, we have 
    \begin{equation}\label{eqn:P_bound_phi}
    P \;\ge\; 1 \;+\; \frac{\pi\,c}{3\,\epsilon^2}\,\frac{2N-1}{\sqrt{\,N-1\,}}
    \;=\; \Theta\!\Big(\frac{\sqrt N}{\epsilon^2}\Big).
    \end{equation}
\end{proposition}
    The proof of Proposition \ref{Prop1} is provided in Appendix \ref{app:Proof00}.
Proposition \ref{Prop1} shows that the dimension of the DCR subspace $P$ is on the order of $\Theta(\frac{\sqrt{N}}{\epsilon^2})$.
Based on the $P$ grid points $\{\widetilde{\phi}_i\}_{i=1}^P$, we construct the DCR subspace matrix
\begin{equation}
    \mathbf{G} = [\mathbf{u}(\widetilde{\phi}_1), \cdots, \mathbf{u}(\widetilde{\phi}_P)] \in\mathbb{C}^{N\times P}.
\end{equation}
With $\mathbf{G}$, $\mathbf{g}(\phi_l)$ can be approximated as
\begin{equation}\label{eqn:g_Qv}
    \mathbf{g}(\phi_l) \approx \mathbf{G}\mathbf{v}_l, ~l = 1, \cdots, L,
\end{equation}
where $\mathbf{v}_l\in\mathbb{C}^{P\times 1}$ denotes the representation coefficient vector.

Proposition \ref{Prop1}  provides a principled approach to construct the DCR subspace matrix
 $\mathbf{G}$. In practice, to represent each $\mathbf{g}(\phi_l)$ accurately with a very low-dimensional subspace, one could first build a dense dictionary $\mathbf{G}$ using a large $P$ and then perform principal component analysis on $\mathbf{G}$ to extract the subspace matrix corresponding to the leading singular values. We adopt this approach in our simulations.

\subsection{Channel Estimation via Atomic Norm Minimization}
\label{subsec:AtomicNorm}
With the low-dimensional representation of the modulated waveforms in (\ref{eqn:g_Qv}), the effective number of degrees of freedom in the near-field channel vector (\ref{eqn:h_BSR}) is reduced to $\Theta(L\sqrt{N})$, which is smaller than the number of samples $N$ when $L\ll \sqrt{N}$.
Therefore, it is feasible to jointly recover the set of complex exponentials parameterized by $\{\alpha_l, \omega_l\}_{l=1}^L$ alongside the modulated waveforms $\{\mathbf{g}(\phi_l)\}_{l=1}^L$.
Following \cite{Yang-2016TIT}, we reformulate near-field channel estimation as a structured low-rank matrix recovery task via a lifting trick and solve it using atomic norm minimization.

Let $\mathbf{q}_n^H$ denote the $n$th row of $\mathbf{G}$ and $\mathbf{e}_n$ be the $n$th column of an identity matrix $\mathbf{I}_N$.
The $n$th element of the channel vector $\mathbf{h}$ can be written as
\begin{equation}\label{eqn:hn2}
    \begin{split}
        [\mathbf{h}]_n &= \sum_{l=1}^L \alpha_l\mathbf{a}^T(\omega_l)\mathbf{e}_n \mathbf{q}_n^H\mathbf{v}_l \\
        & =\text{Tr}\left(\sum_{l=1}^L \alpha_l \mathbf{e}_n \mathbf{q}_n^H \mathbf{v}_l\mathbf{a}^T(\omega_l)\right)\\
        &=\left\langle\sum_{l=1}^L\alpha_l\mathbf{v}_l\mathbf{a}^T(\omega_l), \mathbf{q}_n \mathbf{e}^H_n \right\rangle.
    \end{split}
\end{equation}
Define the matrix $\mathbf{X}_h = \sum_{l=1}^L\alpha_l\mathbf{v}_l\mathbf{a}^T(\omega_l)\in\mathbb{C}^{P\times N}$ and the linear operator $\mathcal{P}: \mathbb{C}^{P\times N} \rightarrow \mathbb{C}^N$ as $[\mathcal{P}(\mathbf{X}_h)]_n = \langle\mathbf{X}_h,\mathbf{q}_n \mathbf{e}^H_n\rangle$.
Then, the near-field channel vector $\mathbf{h}$ is equivalently written as 
\begin{equation}
    \mathbf{h} = \mathcal{P}(\mathbf{X}_h).
\end{equation}
Similarly, defining $M= N_{\text{RF}} Q$ and $\mathcal{P}_{y}:\mathbb{C}^{P\times N}\rightarrow \mathbb{C}^M$ with $[\mathcal{P}_{y}(\mathbf{X})]_m = \sum_n [\mathbf{A}]_{m,n} \langle\mathbf{X},\mathbf{q}_n \mathbf{e}^H_n\rangle$, we  write the received pilot signal $\mathbf{y}$ compactly as 
\begin{equation}\label{eqn:y_operator}
    \mathbf{y} = \mathcal{P}_{y}(\mathbf{X}_h) = \mathbf{A} \mathcal{P}(\mathbf{X}_h).
\end{equation}
By construction, $\mathbf{X}_h$ is a sum of $L$ rank-one matrices, and thus $\rank(\mathbf{X}_h)\leq L$.

In practical scenarios, the number of propagation paths $L$ is small. This motivates us to use the atomic norm to promote sparsity. Define the set of atoms $\mathcal{A}=\{\mathbf{v}\mathbf{a}^T(\omega): \omega\in[0,1), \|\mathbf{v}\|_2=1, \mathbf{v}\in\mathbb{C}^{P \times 1}\}$ and the associated atomic norm as
\begin{equation}
     \begin{split}
          &\|\mathbf{X}\|_{\mathcal{A}}=\inf \{t>0: \mathbf{X}\in t \text{conv}(\mathcal{A})\}\\
          &=\inf_{c_k,\omega_k, \|\mathbf{v}_k\|_2=1} \left\{\sum_{k}|c_k|: \mathbf{X}=\sum_kc_k\mathbf{v}_k\mathbf{a}^T(\omega_k) \right\}.
     \end{split}
\end{equation}
To enforce sparsity of the atomic representation, the channel estimation problem can be solved with the following regularized atomic norm minimization problem

\begin{equation}\label{prob:ANM_Noise}
     \underset{\mathbf{X}}{\text{minimize}} \quad  \frac{1}{2} \|\mathbf{y} - \mathcal{P}_y(\mathbf{X})\|_2^2+\tau\|\mathbf{X}\|_{\mathcal{A}},
\end{equation}
which admits an equivalent SDP \cite{Yang-2016TIT}:
\begin{equation}\label{prob:SDP_Noise}
    \begin{split}
       &  \underset{\mathbf{\mathbf{X},\mathbf{u},\mathbf{V}}}{\text{minimize}} \quad \frac{1}{2} \|\mathbf{y} - \mathcal{P}_y(\mathbf{X})\|_2^2~+ \\
       & ~~~~~~~~~~~~~~~~~~~~~\tau\left(\frac{1}{2N} \text{Tr}(\text{Toep}(\mathbf{u})) +\frac{1}{2}\text{Tr}(\mathbf{V})\right) \\
      & \text{subject to} \quad \begin{bmatrix}
           \text{Toep}(\mathbf{u}) & \mathbf{X}^H\\
           \mathbf{X} & \mathbf{V}
       \end{bmatrix}\succeq 0,\\
    \end{split}
\end{equation}
where $\mathbf{u}$ is a complex vector whose first entry is real, $\text{Toep}(\mathbf{u})$ denotes the $N\times N$ Hermitian Toeplitz
matrix whose first column is $\mathbf{u}$, $\mathbf{V}$ is a $P\times P$ Hermitian matrix, and $\tau>0$ is an appropriately chosen regularization hyperparameter.



The noisy formulation (\ref{prob:SDP_Noise}) can be solved efficiently using off-the-shelf SDP solvers, such as CVX \cite{cvx}.
For large-scale problems, first-order methods, such as the alternating direction method of multipliers (ADMM), can be developed to achieve better computational efficiency.
Denoting the solution to 
(\ref{prob:SDP_Noise}) as $\{\widehat{\mathbf{X}}, \widehat{\mathbf{u}}, \widehat{\mathbf{V}}\}$,
the estimated near-field channel vector is given by $\widehat{\mathbf{h}} = \mathcal{P}(\widehat{\mathbf{X}})$.

\subsection{Recovery Guarantees with Noisy Measurements}
\label{subsec: NoiseCase}
In this subsection, we establish the error bound for  the regularized atomic norm minimization problem in (\ref{prob:ANM_Noise}), in the presence of additive Gaussian noise.
To simplify the analysis, we assume that the noise follows the Gaussian distribution, i.e., $\mathbf{n}\sim \mathcal{CN}(0, \sigma^2\mathbf{I}_M)$, where $\sigma^2$ denotes the noise variance.
Our analysis is inspired by the prior work on atomic norm denoising for line spectral estimation~\cite{Bhaskar-TSP13}. 

Let $\|\cdot\|_{\mathcal{A}}^\ast$ be the dual norm of the atomic norm, which is defined as 
\begin{equation}
    \|\mathbf{Z}\|_{\mathcal{A}}^\ast = \sup_{\| \mathbf{X} \|_{\mathcal{A}}\leq 1} \langle \mathbf{Z}, \mathbf{X}\rangle_{\mathbb{R}} = \sup_{\omega\in[0,1)} \|\mathbf{Z}\overline{\mathbf{a}(\omega)}\|_2.
\end{equation}



\begin{thm}\label{thm1}
    Let $\mathcal{P}^\ast_y: \mathbb{C}^M \mapsto \mathbb{C}^{P\times N}$ be the adjoint operator of $\mathcal{P}_y$. If  
        $\| \mathcal{P}_y^{\ast}(\mathbf{n})\|_\mathcal{A}^{\ast} \le\ \tau$,
    then every minimizer $\widehat{\mathbf{X}}$ of (\ref{prob:ANM_Noise}) satisfies
    \begin{equation}\label{eq:measured-bound}
        \frac{1}{N}\big\|\mathcal{P}_y(\widehat{\mathbf{X}} - \mathbf{X}_h)\big\|_2^2\ \le\ \frac{2 \tau}{N} \|\mathbf{X}_h\|_{\mathcal A}.
    \end{equation}
\end{thm}
    We include the proof of Theorem \ref{thm1} in Appendix \ref{app:Proof1}.

Theorem \ref{thm1} shows that when the regularization hyperparameter $\tau$ is appropriately selected such that 
$\tau \geq \| \mathcal{P}_y^{\ast}(\mathbf{n})\|_\mathcal{A}^{\ast}$, the recovery mean squared error of the low-rank matrix $\mathbf{X}_h$ from the SDP solution $\widehat{\mathbf{X}}$, when projected onto the observed entries, i.e., $\frac{1}{N}\|\mathcal{P}_y(\widehat{\mathbf{X}} - \mathbf{X}_h)\|_2^2$, is upper bounded by $\frac{2\tau}{N}$ times the atomic norm of $\mathbf{X}_h$, which is typically small due to the low-rankness of $\mathbf{X}_h$.


Next, we show that when the noise is i.i.d. Gaussian, the expectation of the dual norm, i.e., $\mathbb{E}(\| \mathcal{P}_y^{\ast}(\mathbf{n})\|_\mathcal{A}^{\ast})$, can be bounded. 

\begin{thm} \label{newthm}
Assume that the entries of $\mathbf{n}$ are i.i.d. Gaussian and $\mathbf{n}\sim\mathcal{CN}(\mathbf{0},\sigma^2\mathbf{I}_M)$. Let $\tau = \sqrt{ 1 + \frac{1}{\log N}} \sqrt{\log N + \log (4\pi \log N) + 1} \|\mathbf{A}\|_2 \|\mathbf{G}\|_F \sigma$. Then we have 
\begin{equation}\label{eq:key_result}
\mathbb{E}(\| \mathcal{P}_y^{\ast}(\mathbf{n})\|_\mathcal{A}^{\ast}) \leq \tau.
\end{equation}
Furthermore,
the minimizer $\widehat{\mathbf{X}}$ of (\ref{prob:ANM_Noise}) satisfies
            \begin{equation}
                \frac{1}{N}\mathbb{E}\,\big\|\mathcal{P}_y(\widehat{\mathbf{X}}-\mathbf{X}_h)\big\|_2^2
                \ \le\ \frac{2\,\tau\,}{N}\|\mathbf{X}_h\|_{\mathcal{A}}.
                \label{eq:new_result}
            \end{equation}
\end{thm}
The proof of the key result in (\ref{eq:key_result}) above is provided in Appendix \ref{app:newthm}. The recovery mean squared error bound in (\ref{eq:new_result}) is obtained immediately by applying Theorem \ref{thm1} and (\ref{eq:key_result}). Note that the error bound (\ref{eq:new_result}) is as tight as the standard error bound results derived in the atomic norm denoising for line spectral estimation problem \cite{Bhaskar-TSP13,Li&Chi-TSP16}. 

Theorem \ref{newthm}  shows that, in the presence of Gaussian noise and when the hyperparameter $\tau$ is chosen according to Theorem \ref{newthm}, solving the regularized atomic norm minimization problem in (\ref{prob:ANM_Noise}) reliably recovers the low-rank matrix $\mathbf{X}_h$ on the observed entries by the measurement operator $\mathcal{P}_y$. 
When the near-field channel is given exactly by $\mathbf{h} = \mathcal{P}(\mathbf{X}_h)$ and denoting its estimate 
 $\widehat{\mathbf{h}} = \mathcal{P}(\widehat{\mathbf{X}})$, it holds that $\frac{1}{N}\mathbb{E} \big\| \mathbf{A}(\widehat{\mathbf{h}} - \mathbf{h}) \big\|_2^2 \leq \frac{2\,\tau\,}{N}\|\mathbf{X}_h\|_{\mathcal{A}} $, which establishes the reliable recovery of the near-field channel $\mathbf{h}$ under the projection of the sensing matrix $\mathbf{A}$. 

\section{Joint Angle and Range Estimation}
\label{sec:ParameterEstimation}
This section develops a two-stage strategy for joint recovery of angle and range parameters from the solution to the convex problem 
(\ref{prob:SDP_Noise}), i.e., $\{\widehat{\mathbf{X}}, \widehat{\mathbf{u}}, \widehat{\mathbf{V}}\}$.
Stage-I extracts gridless angle and
coarse range estimates directly from $\{\widehat{\mathbf{X}}, \widehat{\mathbf{u}}, \widehat{\mathbf{V}}\}$.
Stage-II refines both angle and range estimates under the exact spherical-wave steering model via gradient-based methods.

\subsection{Gridless Angle and Coarse Range Estimation}
\label{subsec:ParameterEstimation}
\paragraph*{1) {\bf Gridless Angle Estimation}}
Gridless angle estimation can be obtained by performing Vandermonde decomposition on the Toeplitz matrix $\text{Toep}(\widehat{\mathbf{u}})$, yielding $L$ frequencies $\{\widehat{\omega}_l\}_{l=1}^L$.
Since $\omega_l = \frac{d\sin{\theta_l}}{\lambda_c}$, the corresponding angle estimate is given by
\begin{equation}\label{eqn:theta_l}
    \widehat{\theta}_l = \arcsin{\frac{\widehat{\omega}_l\lambda_c}{d}},\quad l=1,\cdots,L.
\end{equation}

\paragraph*{2) {\bf Coarse Range Estimation}}
Given the estimates $\{\widehat{\omega}_l\}_{l=1}^L$, we substitute them into (\ref{eqn:hn2}). Then (\ref{eqn:y_operator}) reduces to the following overdetermined linear system of equations:
 \begin{equation}\label{eqn:LS_eqn}
        \mathbf{A}\begin{bmatrix}
            \mathbf{a}^T(\widehat{\omega}_1)\mathbf{e}_1\mathbf{q}_1^H &\cdots &\mathbf{a}^T(\widehat{\omega}_L)\mathbf{e}_1\mathbf{q}_1^H\\
            \vdots & \ddots & \vdots\\
            \mathbf{a}^T(\widehat{\omega}_1)\mathbf{e}_N\mathbf{q}_N^H&\cdots&\mathbf{a}^T(\widehat{\omega}_L)\mathbf{e}_N\mathbf{q}_N^H
        \end{bmatrix}\begin{bmatrix}
            \alpha_1 \mathbf{v}_1\\
            \vdots\\
            \alpha_L\mathbf{v}_L
        \end{bmatrix}=\mathbf{y}
\end{equation}
An estimate of the vector $[\alpha_1 \mathbf{v}_1^T, \cdots, \alpha_L \mathbf{v}_L^T]^T$ is obtained via least squares.
Note that we cannot resolve the intrinsic scaling ambiguity between each $\alpha_l$ and the corresponding $\mathbf{v}_l$ due to their multiplication.

We perform coarse range estimation by exploiting the structure of $\mathbf{g}(\phi_l)$.
According to (\ref{eqn:h_BSR}) and (\ref{eqn:g_Qv}), we have $\alpha_l\mathbf{g}(\phi_l) = \alpha_l\mathbf{G} \mathbf{v}_l$.
Let $\boldsymbol{\rho}_l^{(1)}= \left[[\alpha_l\mathbf{g}(\phi_l)]_0,\cdots, [\alpha_l\mathbf{g}(\phi_l)]_{N-2}\right]^T$ and $\boldsymbol{\rho}_l^{(2)}= \left[[\alpha_l\mathbf{g}(\phi_l)]_1,\cdots, [\alpha_l\mathbf{g}(\phi_l)]_{N-1}\right]^T$.
To eliminate the effect of $\alpha_l$, we construct the vector $\boldsymbol{\rho}_l =\boldsymbol{\rho}_l^{(2)} \oslash \boldsymbol{\rho}_l^{(1)}
    =[e^{j2\pi\phi_l}, e^{j6\pi\phi_l},\cdots, e^{j2(2N-3)\pi\phi_l}]^T$,
where $\oslash$ denotes element-wise division.
Since $\phi_l = \frac{d^2 \cos^2 \theta_l}{2\lambda_c r_l}$, we have
\begin{equation}\label{eqn:r_l}
    r_l = \frac{j\pi d^2 \cos^2{\theta_l}}{\lambda_c \mathbf{n}_0^\dagger \log(\boldsymbol{\rho}_l)},
\end{equation}
where $\mathbf{n}_0 = [1,3,\cdots,2N-3]^T$ and $(\cdot)^\dagger$ denotes the  Moore–Penrose inverse.
Finally, substituting the estimates $\widehat{\theta}_l$ and $\widehat{\mathbf{\boldsymbol{\rho}}}_{l}$ into (\ref{eqn:r_l}) yields the coarse range estimates $\{\widehat{r}_l\}_{l=1}^L$.

\subsection{Refinement under the Exact Spherical Model}
\label{subsec:RangeRefinement}
To mitigate the modeling bias introduced by the second-order approximation in (\ref{eqn:rn_approx}), we refine both the angle and range estimates using the exact spherical-wave steering vector (\ref{eqn:b_theta_r}).
Starting from the initial estimates $\{\widehat{\theta}_l\}_{l=1}^L$ and $\{\widehat{r}_l\}_{l=1}^L$ derived in Subsection \ref{subsec:ParameterEstimation}, we adopt an alternating refinement approach \cite{Cui-ToC2022} that updates the angle and range estimates sequentially while eliminating the effect of path gains via orthogonal projection.

With the exact near-field steering vector $\mathbf{b}(\theta_l, r_l)$ in (\ref{eqn:b_theta_r}), define the following matrices
\begin{equation}
\begin{cases}
    \mathbf{W}(\boldsymbol{\theta},\mathbf{r}) = [\mathbf{b}(\theta_1, r_1), \cdots, \mathbf{b}(\theta_L, r_L)]\in\mathbb{C}^{N\times L},\\
    \mathbf{S}(\boldsymbol{\theta},\mathbf{r}) = \mathbf{A}\mathbf{W}(\boldsymbol{\theta},\mathbf{r}) \in\mathbb{C}^{M\times L},
\end{cases}
\end{equation}
where $\boldsymbol{\theta}= [\theta_1, \cdots, \theta_L]^T$ and $\boldsymbol{r}= [r_1, \cdots, r_L]^T$. 
The near-field channel vector satisfies $\mathbf{h} = \mathbf{W}(\boldsymbol{\theta},\mathbf{r}) \boldsymbol{\alpha}$, where $\boldsymbol{\alpha}= [\alpha_1, \cdots, \alpha_L]^T$ denotes the path gain.
We solve the following nonlinear least squares problem
\begin{equation} \label{prob:nls}
    \text{minimize}_{\boldsymbol{\theta},\mathbf{r},\boldsymbol{\alpha}} \|\mathbf{y} -  \mathbf{S}(\boldsymbol{\theta},\mathbf{r}) \boldsymbol{\alpha}\|_2^2,
\end{equation}
which is nonconvex due to the angle-range coupling.
For a fixed pair of $(\boldsymbol{\theta},\mathbf{r})$, the optimal gain vector is $\widehat{\boldsymbol{\alpha}} = \mathbf{S}(\boldsymbol{\theta},\mathbf{r})^\dagger \mathbf{y}$.
Eliminating $\boldsymbol{\alpha}$ yields the following cost function
\begin{equation}\label{eqn: L_cost}
    \mathcal{L}(\boldsymbol{\theta},\mathbf{r}) = \min_{\boldsymbol{\alpha}} \|\mathbf{y} -  \mathbf{S}(\boldsymbol{\theta},\mathbf{r}) \boldsymbol{\alpha}\|_2^2 = \|\boldsymbol{\Pi}_{\mathbf{S}}^\perp(\boldsymbol{\theta},\mathbf{r})\mathbf{y}\|_2^2,
\end{equation}
where the orthogonal projection is defined as $\boldsymbol{\Pi}_{\mathbf{S}}^\perp(\boldsymbol{\theta},\mathbf{r}) = \mathbf{I}_M -\mathbf{S}(\boldsymbol{\theta},\mathbf{r})\mathbf{S}(\boldsymbol{\theta},\mathbf{r})^\dagger$.
Thus, solving (\ref{prob:nls}) reduces to
\begin{equation}\label{eqn: L_cost_red}
    \min_{\boldsymbol{\theta},\mathbf{r}} \mathcal{L}(\boldsymbol{\theta},\mathbf{r}).
\end{equation}
The optimization problem (\ref{eqn: L_cost_red}) can be solved using an iterative gradient descent approach, starting from initial estimates $\boldsymbol{\theta}^0 = \widehat{\boldsymbol{\theta}}$ and $\mathbf{r}^0 = \widehat{\mathbf{r}}$.
In the $n$th iteration, the angle and range estimates are updated sequentially as
\begin{equation}
\begin{cases}
    \boldsymbol{\theta}^n = \boldsymbol{\theta}^{n-1} - \xi_1 \nabla_{\boldsymbol{\theta}}  \mathcal{L}(\boldsymbol{\theta},\mathbf{r}^{n-1})|_{\boldsymbol{\theta} = \boldsymbol{\theta}^{n-1}} \\
    \mathbf{r}^n = \mathbf{r}^{n-1} - \xi_2 \nabla_{\mathbf{r}}  \mathcal{L}(\boldsymbol{\theta}^n,\mathbf{r})|_{\mathbf{r} = \mathbf{r}^{n-1}},
    \end{cases}
\end{equation}
where $\xi_1,\xi_2>0$ denote the step sizes.
To ensure the objective function is non-increasing, the step sizes $\xi_1$ and $\xi_2$ are chosen via Armijo backtracking line search.

The gridless angle estimates $\{\widehat{\theta}_l\}_{l=1}^L$ and coarse range estimates $\{\widehat{r}_l\}_{l=1}^L$ obtained in Subsection \ref{subsec:ParameterEstimation} provide a superior initialization to facilitate the convergence of the iterative gradient descent refinement.
The proposed  two-stage algorithm is summarized in Algorithm \ref{alg}.

\begin{algorithm}[t!]
        \caption{The Proposed Super-Resolution Near-Field Channel Estimation Algorithm.}
        \label{alg}
        \begin{algorithmic}[1]
            \REQUIRE $\mathbf{y}$, $\mathbf{A}$, $\mathbf{G}$, $N$, $L$, $d$, $\lambda_c$, $\xi_1$, $\xi_2$, and $N_{\text{iter}}$\\         
            \ENSURE $\widehat{\boldsymbol{\theta}}$, $\widehat{\mathbf{r}}$, $\widehat{\boldsymbol{\alpha}}$, $\widehat{\mathbf{h}}$. \\
            \textbf{Stage~I: Initial Estimation via ANM}
            
            \STATE Solve the SDP 
            \eqref{prob:SDP_Noise} to obtain 
            $(\widehat{\mathbf{X}},\widehat{\mathbf{u}},\widehat{\mathbf{V}})$.\\
            
            \STATE Obtain $L$ frequencies $\{\widehat{\omega}_l\}_{l=1}^L$ by performing Vandermonde decomposition of $\mathrm{Toep}(\widehat{\mathbf{u}})$. 
            
            \STATE Convert $\{\widehat{\omega}_l\}_{l=1}^L$ to $\{\widehat{\theta}_l\}_{l=1}^L$ using \eqref{eqn:theta_l}.

            \STATE Substitute $\{\widehat{\omega}_l\}_{l=1}^L$ into \eqref{eqn:LS_eqn} and solve the least squares problem to obtain the estimate of the vector $[\alpha_1 \mathbf{v}_1^T, \cdots, \alpha_L \mathbf{v}_L^T]^T$.

            \STATE For each $l$, obtain $\widehat{r}_l$ by using \eqref{eqn:r_l}.\\

            \textbf{Stage~II: Refinement Estimation via Gradient Descent}
            
            \STATE Define $\mathbf{W}(\boldsymbol{\theta},\mathbf{r})$ and $\mathbf{S}(\boldsymbol{\theta},\mathbf{r})$, and form the cost function $\mathcal{L}(\boldsymbol{\theta},\mathbf{r})=\|\boldsymbol{\Pi}^{\perp}_{\mathbf{S}}(\boldsymbol{\theta},\mathbf{r})\mathbf{y}\|_2^2$ as in \eqref{eqn: L_cost}.
            
            \STATE Initialize  $\boldsymbol{\theta}^0 = \{\widehat{\theta}_l\}_{l=1}^L$ and $\mathbf{r}^0 = \{\widehat{r}_l\}_{l=1}^L$.
            \FOR{$ n= 1,2,\cdots, N_{\text{iter}}$}
            
            \STATE Update angles: $\boldsymbol{\theta}^{n} =
            \boldsymbol{\theta}^{n-1} - \xi_1\,\nabla_{\boldsymbol{\theta}}
                \mathcal{L}(\boldsymbol{\theta},\mathbf{r}^{n-1})\big|_{\boldsymbol{\theta}=\boldsymbol{\theta}^{n-1}}$.
                
            \STATE Update ranges: $\mathbf{r}^{n} = \mathbf{r}^{n-1} - \xi_2\,\nabla_{\mathbf{r}}
                \mathcal{L}(\boldsymbol{\theta}^{n},\mathbf{r})\big|_{\mathbf{r}=\mathbf{r}^{n-1}}$.
            \IF{Convergence}
            \STATE Break;
            \ELSE
            \STATE Continue.
            \ENDIF
            \ENDFOR
            \STATE Update the angles $\widehat{\boldsymbol{\theta}} = \boldsymbol{\theta}^n$ and ranges $\widehat{\mathbf{r}} = \mathbf{r}^n$.
            \STATE Update the path gain $\widehat{\boldsymbol{\alpha}}= \mathbf{S}(\widehat{\boldsymbol{\theta}},\widehat{\mathbf{r}})^\dagger \mathbf{y}$
            \STATE Update the channel vector $\widehat{\mathbf{h}} = \mathbf{W}(\widehat{\boldsymbol{\theta}},\widehat{\mathbf{r}})\widehat{\boldsymbol{\alpha}}$.
        \end{algorithmic}        
    \end{algorithm}

\begin{figure*}[t]
        \centering
        \begin{subfigure}[b]{0.6\columnwidth}
        \includegraphics[width=1\linewidth]{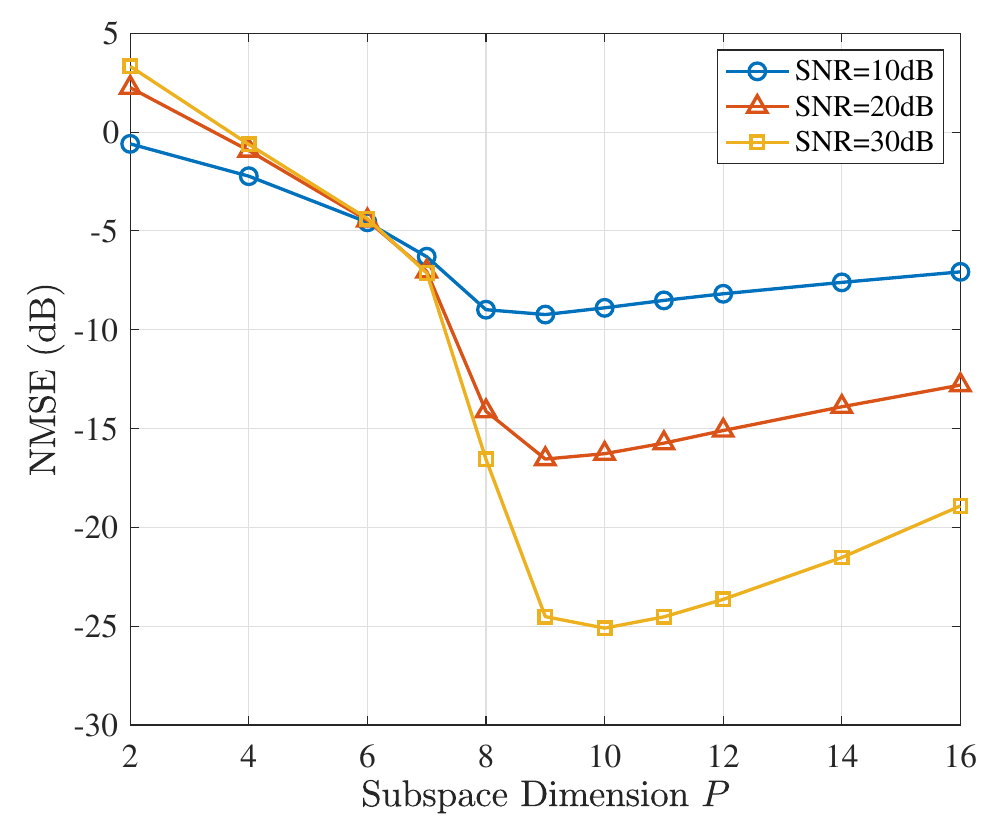}
        \caption{NMSE of $\mathbf{h}$.}   
        \end{subfigure}
        \begin{subfigure}[b]{0.6\columnwidth}
            \includegraphics[width=1\linewidth]{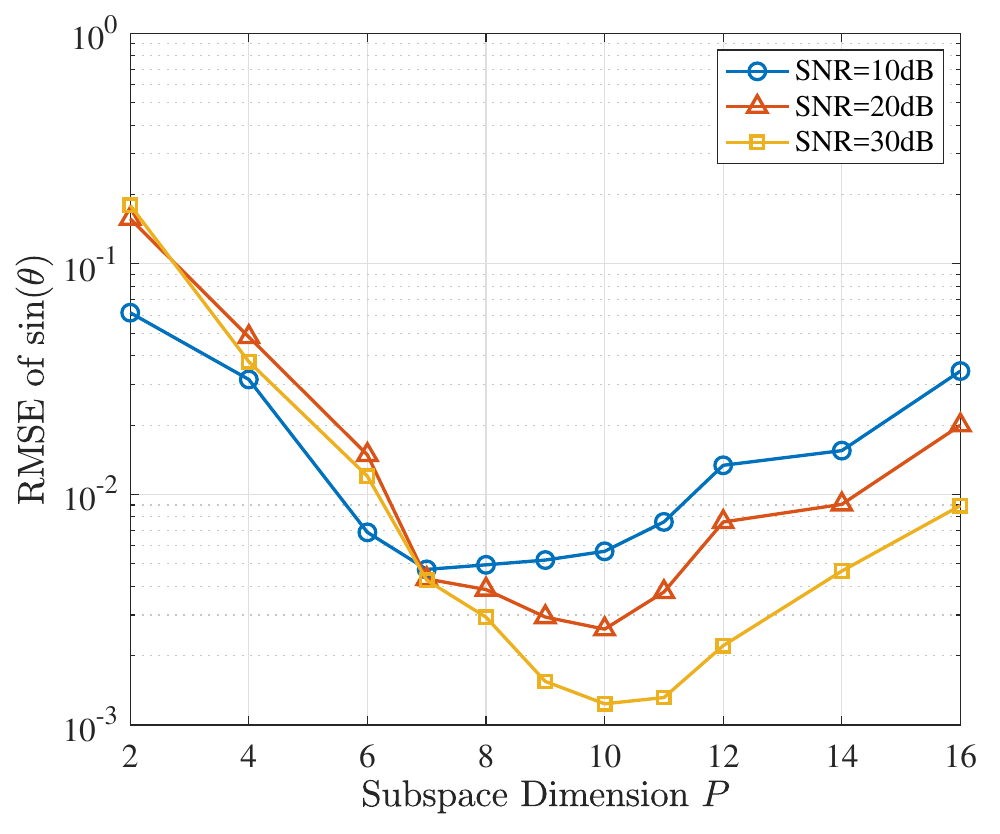}
        \caption{RMSE of $\sin(\theta)$.}  
        \end{subfigure}
        \begin{subfigure}[b]{0.6\columnwidth}
            \includegraphics[width=1\linewidth]{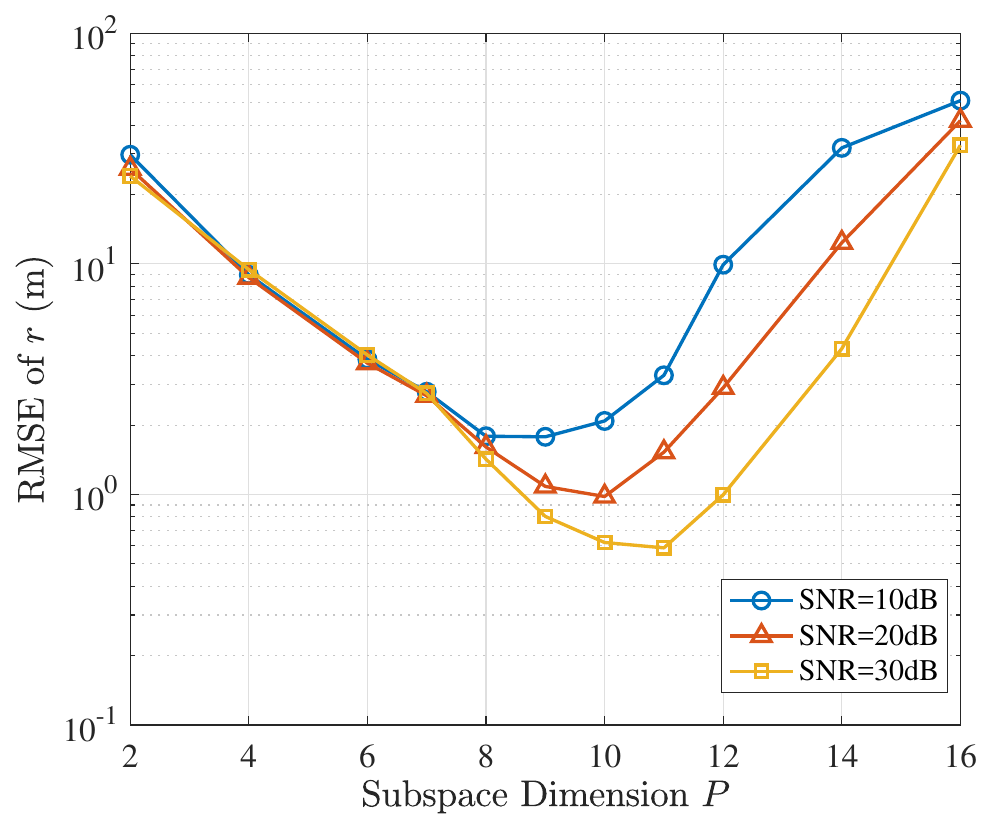}
        \caption{RMSE of $r$.}  
        \end{subfigure}
        \caption{Performance of the ANM(SI) algorithm with respect to different DCR subspace dimensions $P$, where $r\in [5,10]$m.}
        \label{fig:nmse_vs_rank}
\end{figure*} 
\subsection{Computational Complexity}
\label{subsec:complexity}

In this subsection, we analyze the computational complexity of the proposed method in 
Algorithm \ref{alg}. 
Since Stage-I requires solving the SDP in
\eqref{prob:SDP_Noise}, its cost dominates that of Stage-II. Therefore, we focus solely on the complexity of Stage-I.
 According to \cite{Boyd-SDP-SIAM}, when an SDP is solved using interior-point methods (e.g., CVX/SDPT3), the computational complexity is
$\mathcal{O}((N_1)^{2}(N_2)^{2.5})$, where $N_1$ is the number of optimization variables and $N_2 \times N_2$ is the size of the PSD matrix. Therefore, the computational complexity for solving (\ref{prob:SDP_Noise}) is $\mathcal{O}((NP)^{2}(N + P)^{2.5})$. When first‑order solvers (e.g., ADMM) are applied to solve the SDP, the computational complexity can be reduced to $\mathcal{O}(K_{\rm 1st}(N+P)^3)$, where $K_{\rm 1st}$ is the number of first-order iterations. Therefore, first-order methods significantly reduce the computational complexity for solving (\ref{prob:SDP_Noise}). We leave the development of fast solvers, such as ADMM, to future work. 

\section{Numerical Simulations}
\label{sec:Sims}



In this section, we present simulation results to validate the effectiveness of the proposed near-field channel estimation scheme. To evaluate the contribution of each stage in Algorithm~\ref{alg}, we consider two variants of our approach: (i) a Stage-I-only method, denoted as ``ANM(SI)'', and (ii) a two-stage method that employs both Stage-I and Stage-II steps, denoted as ``ANM(SI+SII)''. 
We compare these methods with four representative baselines: the on-grid angular-domain SW-OMP~\cite{Rodriguez-2018TWC}, the on-grid polar-domain P-SOMP\footnote{Note that the original source code of the P-SOMP algorithm sets the number of propagation paths parameter to $2L$ (i.e., double the actual number of paths). For consistent comparisons, we set this parameter to $L$ in our simulations.}~\cite{Cui-ToC2022}, the off-grid angular-domain SS-SIGW-OLS \cite{Gonzalez-TWC2021}, and the off-grid polar-domain P-SIGW \cite{Cui-ToC2022}. 

The simulations consider a MIMO communication system in which the BS is equipped with $N=256$ antennas, operates at a carrier frequency of $f_c = 100\text{GHz}$, and uses $N_{\text{RF}} = 4$ RF chains. The corresponding Rayleigh distance for this configuration is approximately $98\text{m}$. 
The user transmits a pilot sequence of length $Q=32$.
The near-field channel is assumed to consist of $L=4$ propagation paths, including one LOS path and three NLOS paths. The dimension of the DCR subspace is set to $P=10$.
The regularization hyperparameter $\tau$ is selected according to Theorem \ref{newthm}. 
Unless otherwise specified,  the simulations use the parameter configuration described above. 

To quantitatively evaluate the channel estimation accuracy, we adopt the normalized mean squared error (NMSE), defined as $\text{NMSE}=\mathbb{E}\left(\frac{\|\widehat{\mathbf{h}}-\mathbf{h}\|^2_2}{\|\mathbf{h}\|_2^2}\right)$, where $\mathbf{h}$ is the true near-field channel and $\widehat{\mathbf{h}}$ is its estimate.
To assess the estimation accuracy of angle and range parameters, we use the root mean squared errors (RMSEs) of $\sin(\theta)$ and $r$, defined as $\text{RMSE}_{\sin(\theta)} =\sqrt{\frac{1}{L}\sum_{l=1}^L (\sin(\widehat\theta_l) - \sin(\theta_l))^2}$ and $\text{RMSE}_r=\sqrt{\frac{1}{L}\sum_{l=1}^L (\widehat r_l - r_l)^2}$, respectively.
\color{black}


\begin{figure*}[t]
        \centering
        \begin{subfigure}[b]{0.6\columnwidth}
        \includegraphics[width=1\linewidth]{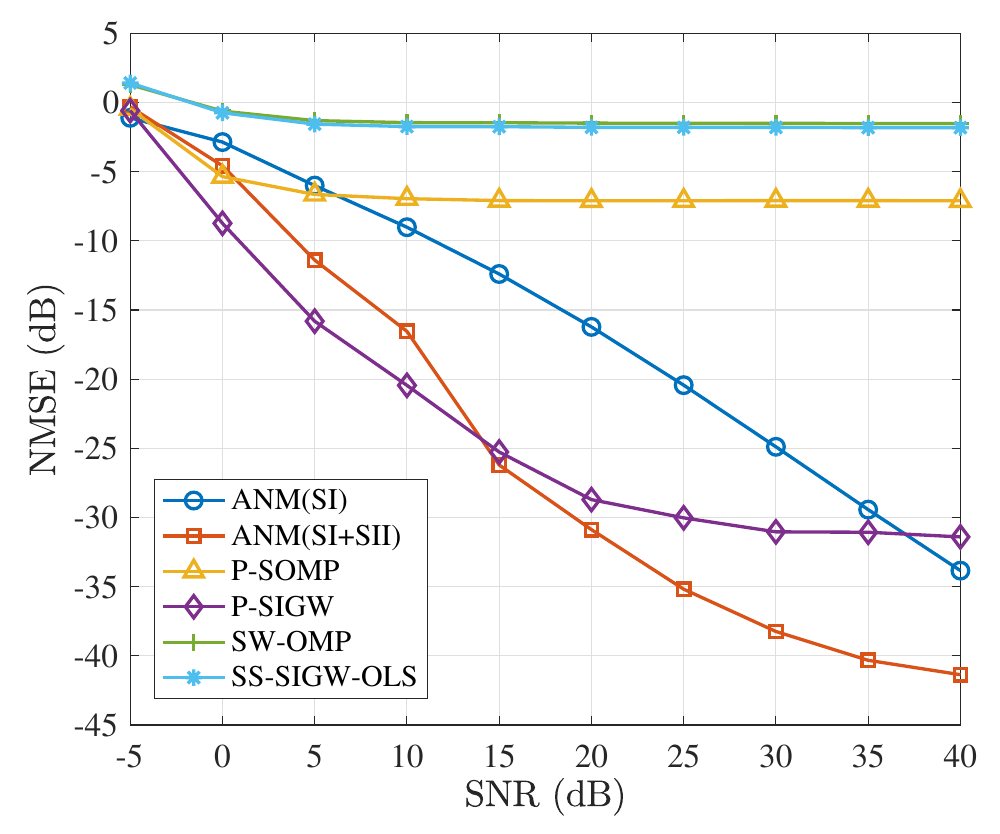}
        \caption{NMSE of $\mathbf{h}$ with $r\in[5,10]\text{m}$.}   
        \end{subfigure}
        \begin{subfigure}[b]{0.6\columnwidth}
            \includegraphics[width=1\linewidth]{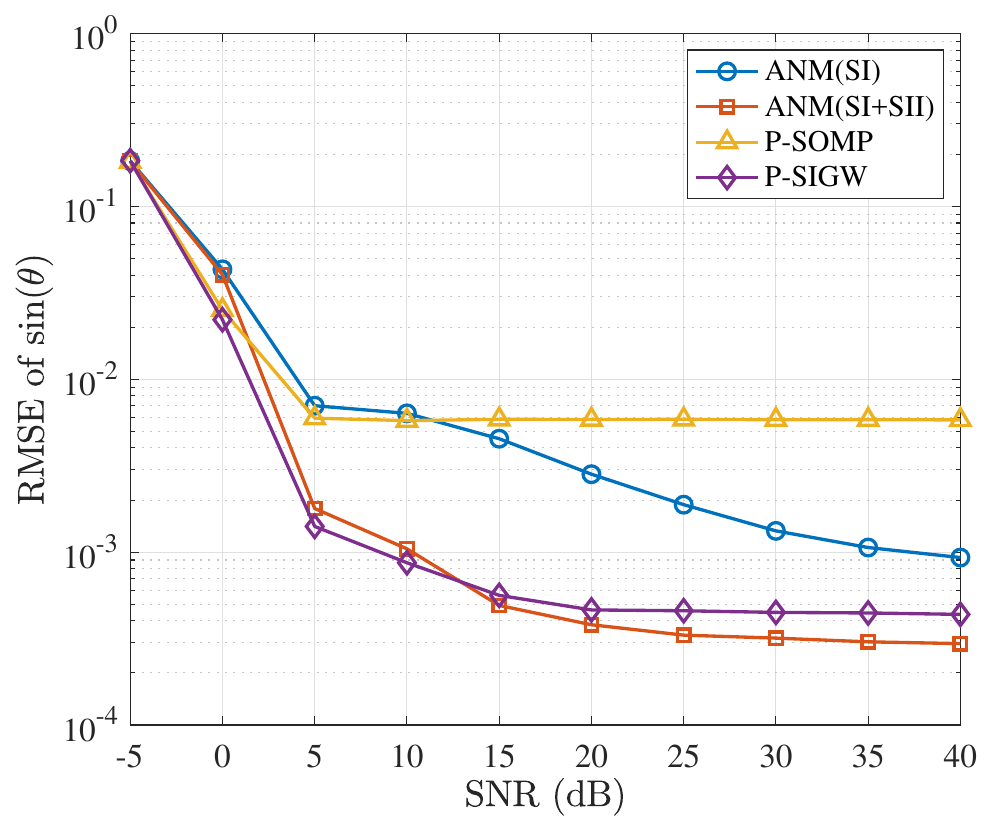}
        \caption{RMSE of $\sin(\theta)$ with $r\in[5,10]\text{m}$.}  
        \end{subfigure}
        \begin{subfigure}[b]{0.6\columnwidth}
            \includegraphics[width=1\linewidth]{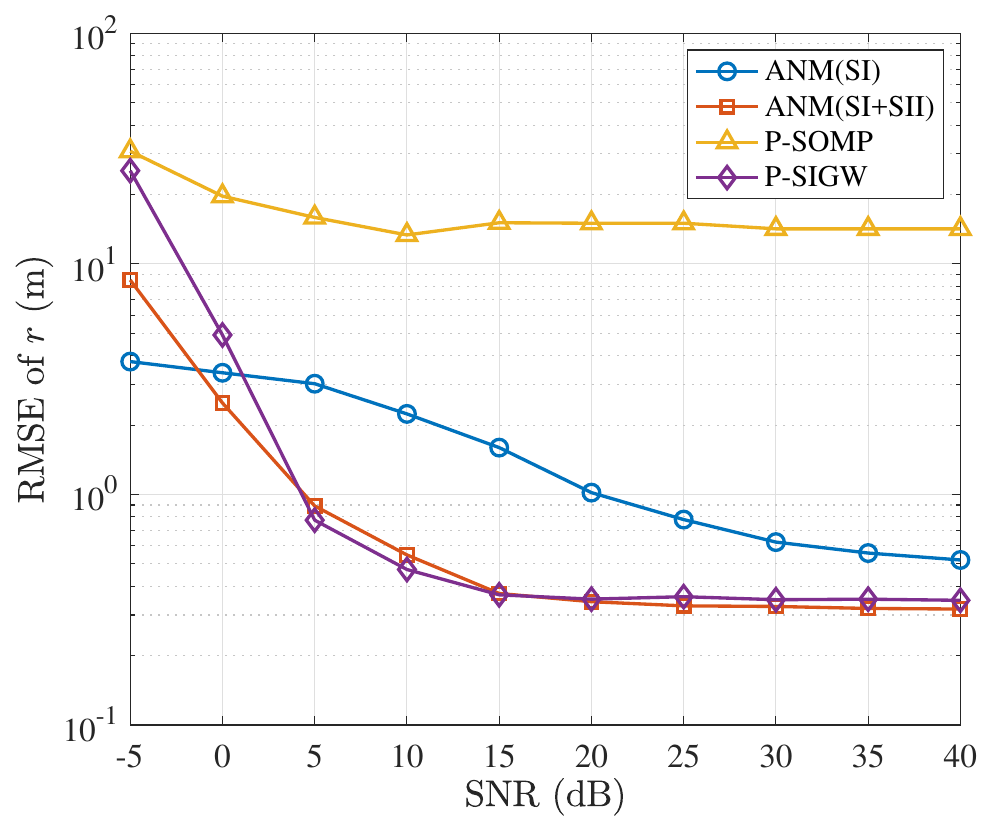}
        \caption{RMSE of $r$ with $r\in[5,10]\text{m}$.}  
        \end{subfigure}\\
        \begin{subfigure}[b]{0.6\columnwidth}
        \includegraphics[width=1\linewidth]{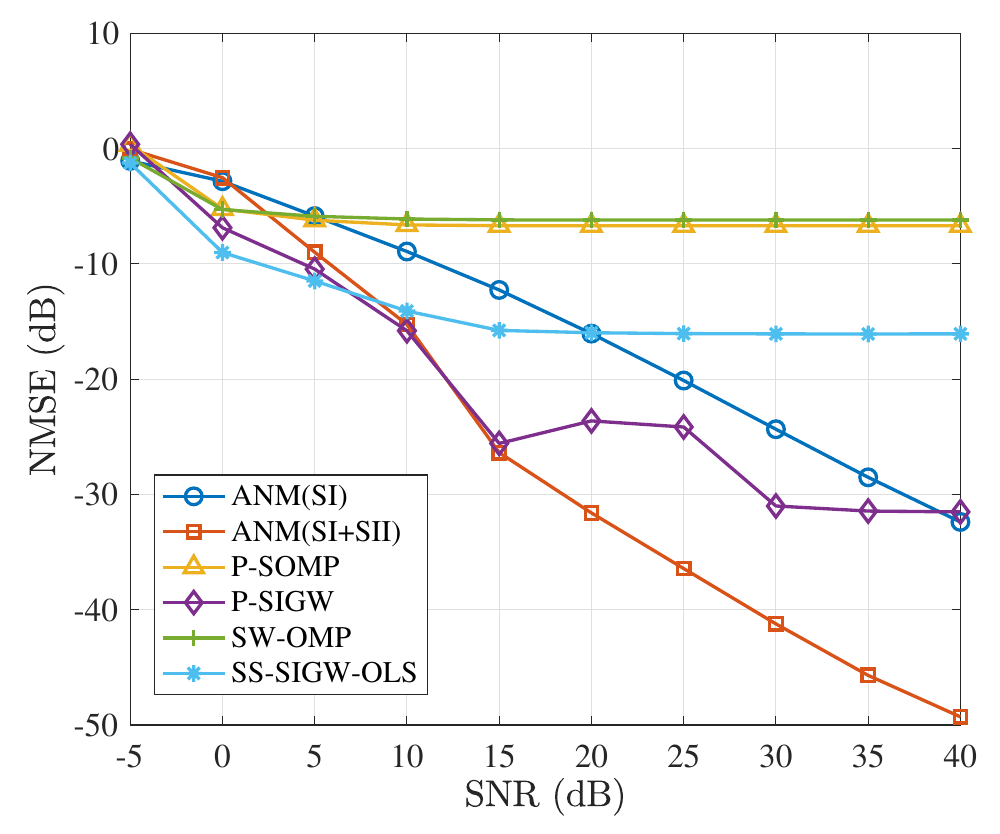}
        \caption{NMSE of $\mathbf{h}$ with $r\in[60,80]\text{m}$.}   
        \end{subfigure}
        \begin{subfigure}[b]{0.6\columnwidth}
            \includegraphics[width=1\linewidth]{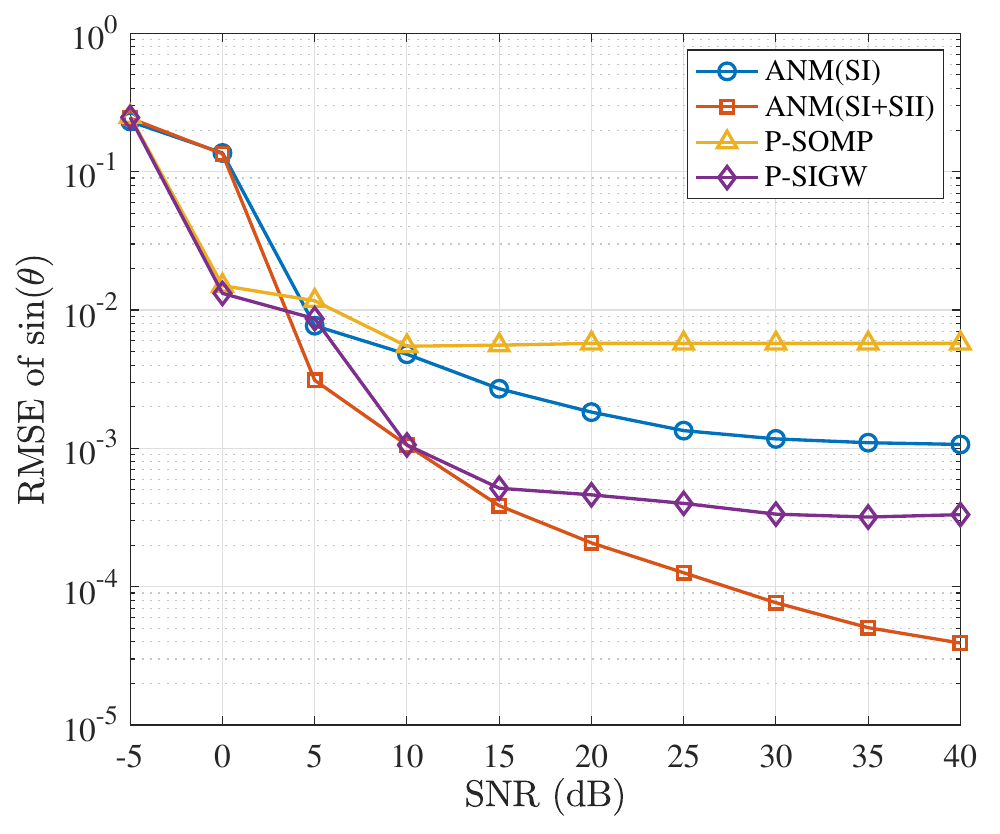}
        \caption{RMSE of $\sin(\theta)$ with $r\in[60,80]\text{m}$.}  
        \end{subfigure}
        \begin{subfigure}[b]{0.6\columnwidth}
            \includegraphics[width=1\linewidth]{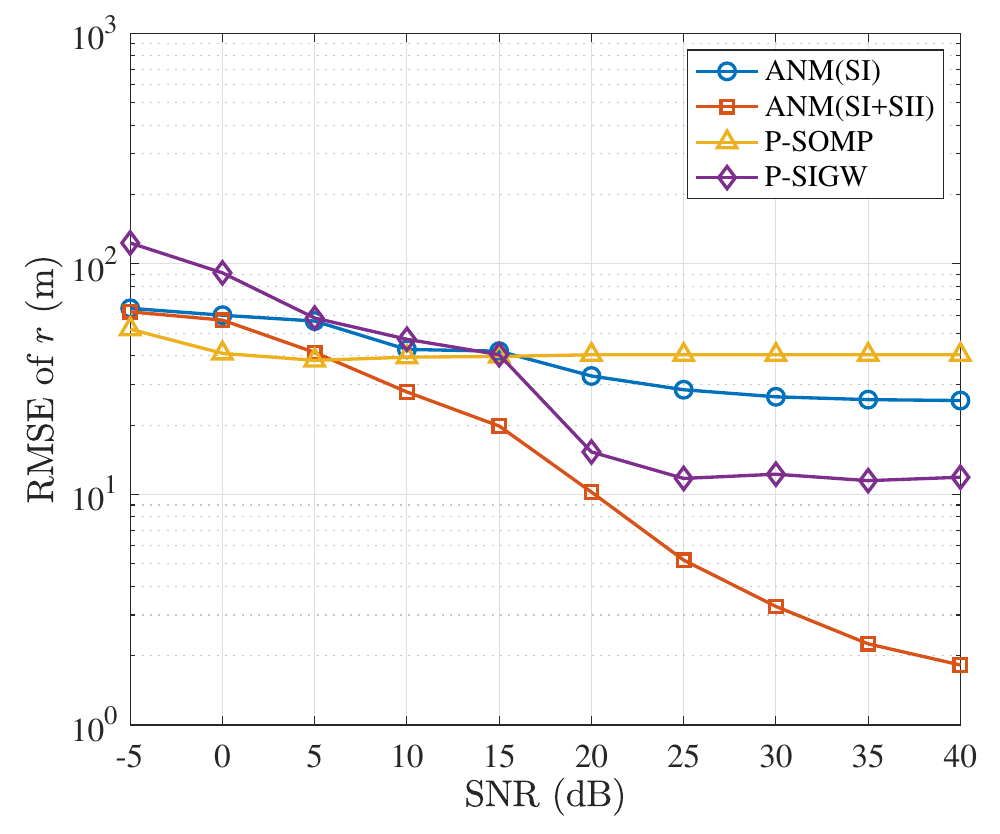}
        \caption{RMSE of $r$ with $r\in[60,80]\text{m}$.}  
        \end{subfigure}
        \caption{Comparison of NMSE versus SNR for the proposed approaches and competing baselines.}
        \label{fig:nmse_vs_SNR}
\end{figure*}  

\subsection{Selection of DCR Subspace Dimension}
\label{subsec:DCRdim}
To determine an appropriate value for the DCR subspace dimension $P$, we evaluate the performance of ANM(SI) for different values of $P$. In this study, the ranges between the BS and the users or scatterers are uniformly drawn from $\mathcal{U}(5, 10)$ meters.

As shown in Fig. \ref{fig:nmse_vs_rank}(a), selecting $P$ around $9$ or $10$ yields the lowest NMSE across SNR levels of 10, 20, and 30 dB.
 This indicates that a moderate subspace dimension sufficiently captures the near-field structure while avoiding excessive model complexity. Fig. \ref{fig:nmse_vs_rank}(b) shows that the angle estimation RMSE reaches its minimum near $P=10$ at moderate and high SNRs, and near $P=7$ at the SNR level of 10 dB. 
Similar trends are observed for range estimation in Fig. \ref{fig:nmse_vs_rank}(c), where comparable $P$ values achieve the best RMSE.
These results demonstrate that an appropriate DCR subspace dimension effectively characterizes the spherical wavefront curvature and enables reliable angle and range recovery for our proposed approach. In all subsequent simulations, we set $P=10$, which achieves an effective balance between estimation accuracy and computational efficiency.

\begin{figure*}[t]
        \centering
        \begin{subfigure}[b]{0.6\columnwidth}
        \includegraphics[width=1\linewidth]{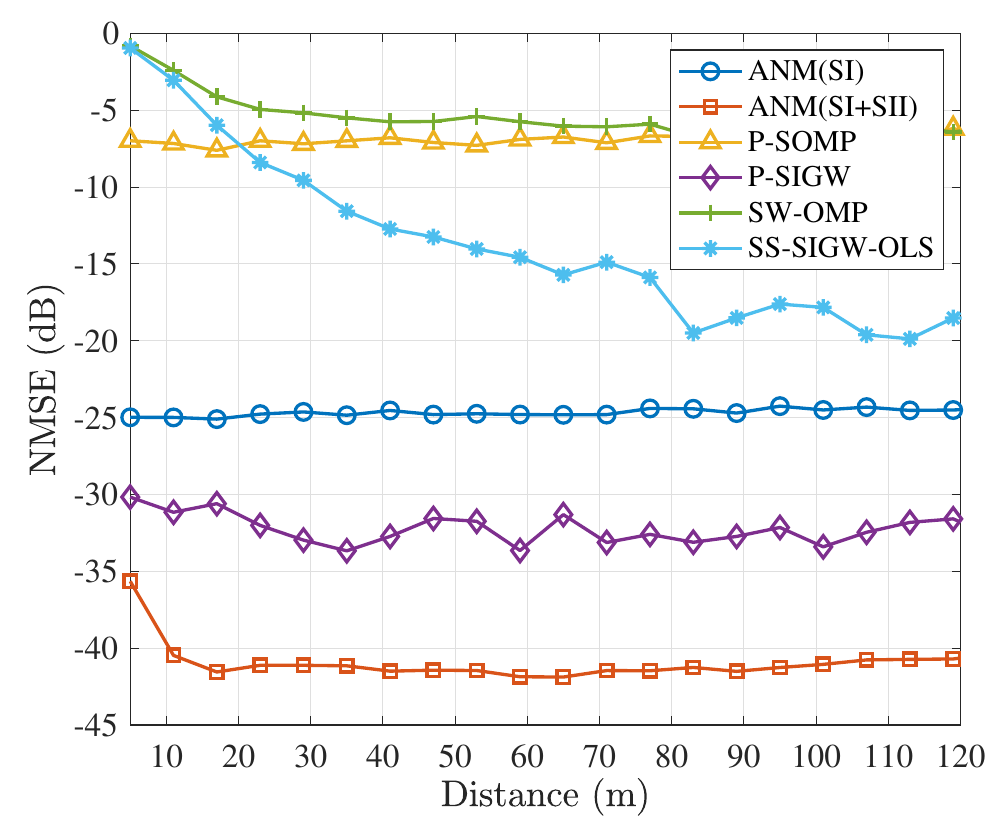}
        \caption{NMSE of $\mathbf{h}$.}   
        \end{subfigure}
        \begin{subfigure}[b]{0.6\columnwidth}
            \includegraphics[width=1\linewidth]{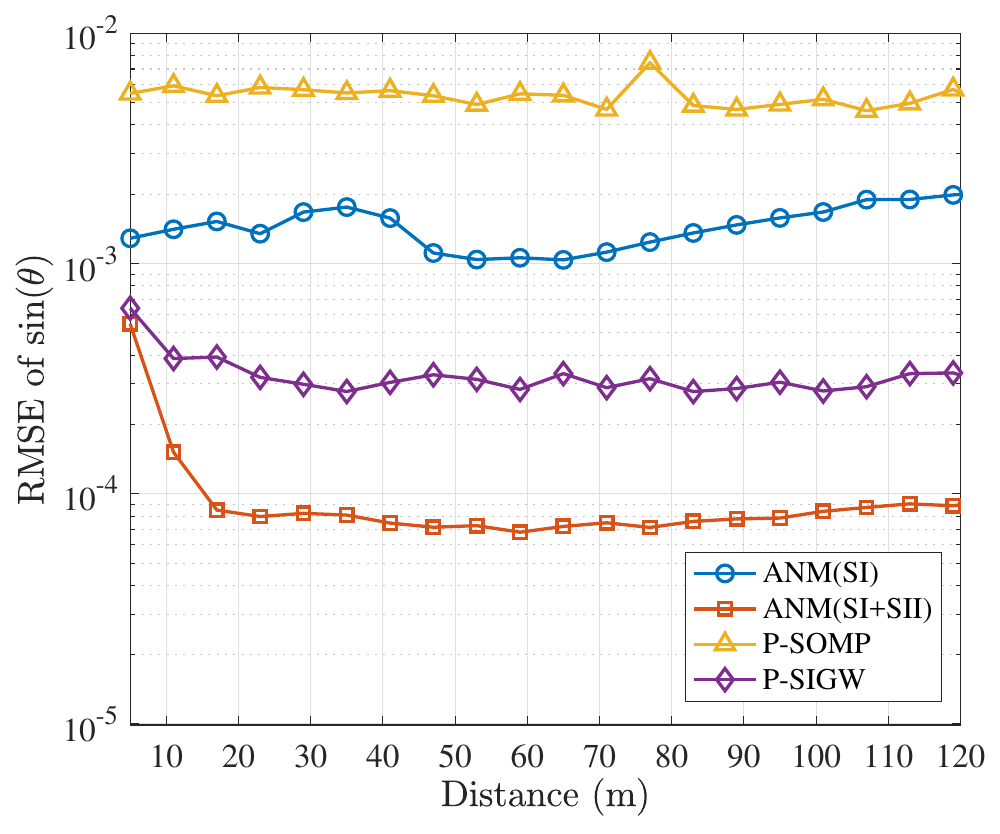}
        \caption{RMSE of $\sin(\theta)$.}  
        \end{subfigure}
        \begin{subfigure}[b]{0.6\columnwidth}
            \includegraphics[width=1\linewidth]{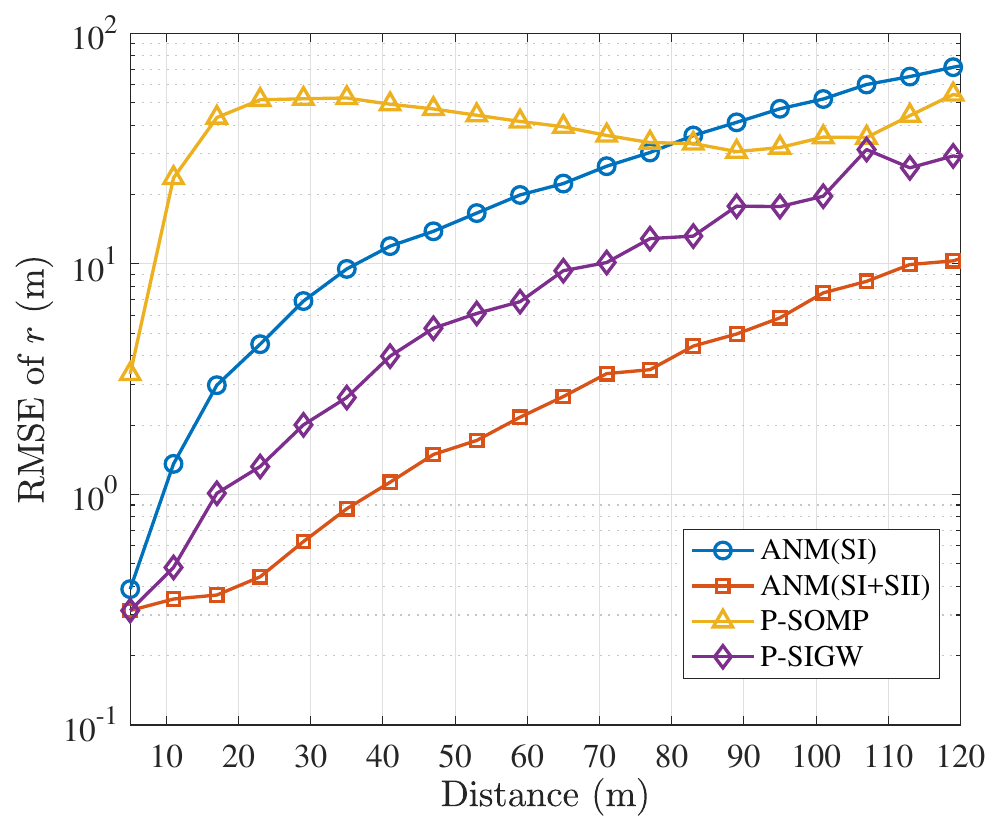}
        \caption{RMSE of $r$.}  
        \end{subfigure}
        \caption{NMSE performance of the proposed approaches and competing baselines versus the distances between BS and users or scatterers when SNR = $30$ dB.}
        \label{fig:nmse_vs_Distance}
\end{figure*} 

\begin{figure*}[t]
        \centering
        \begin{subfigure}[b]{0.6\columnwidth}
        \includegraphics[width=1\linewidth]{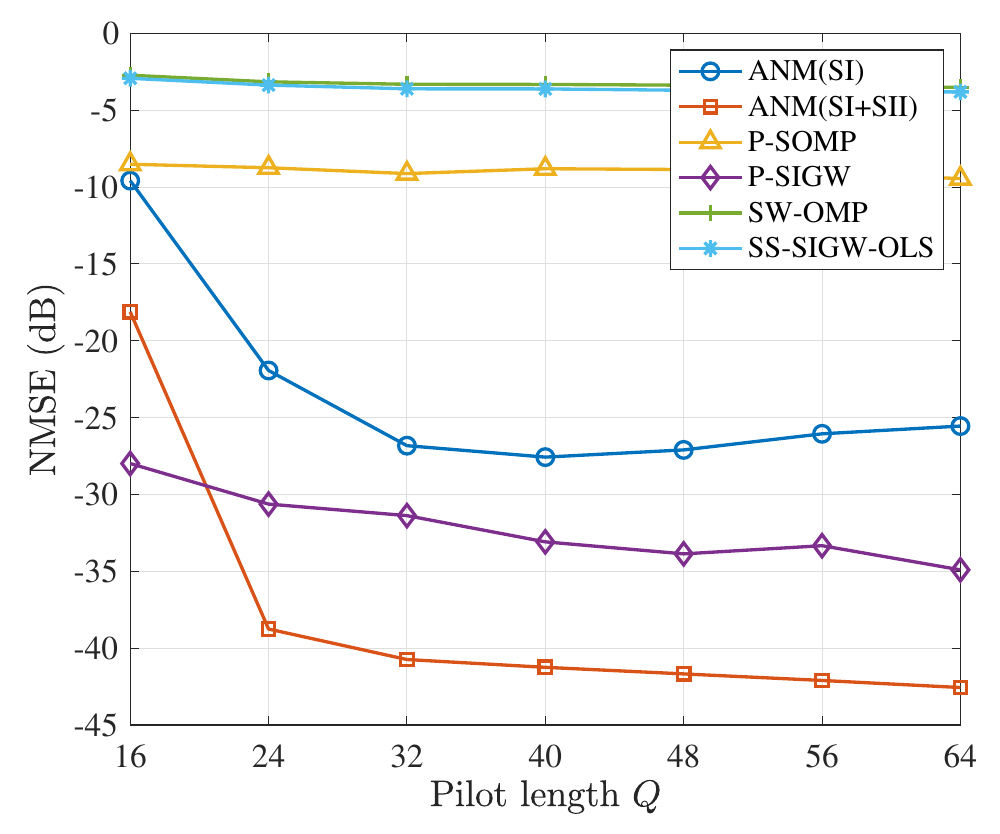}
        \caption{NMSE of $\mathbf{h}$.}   
        \end{subfigure}
        \begin{subfigure}[b]{0.6\columnwidth}
            \includegraphics[width=1\linewidth]{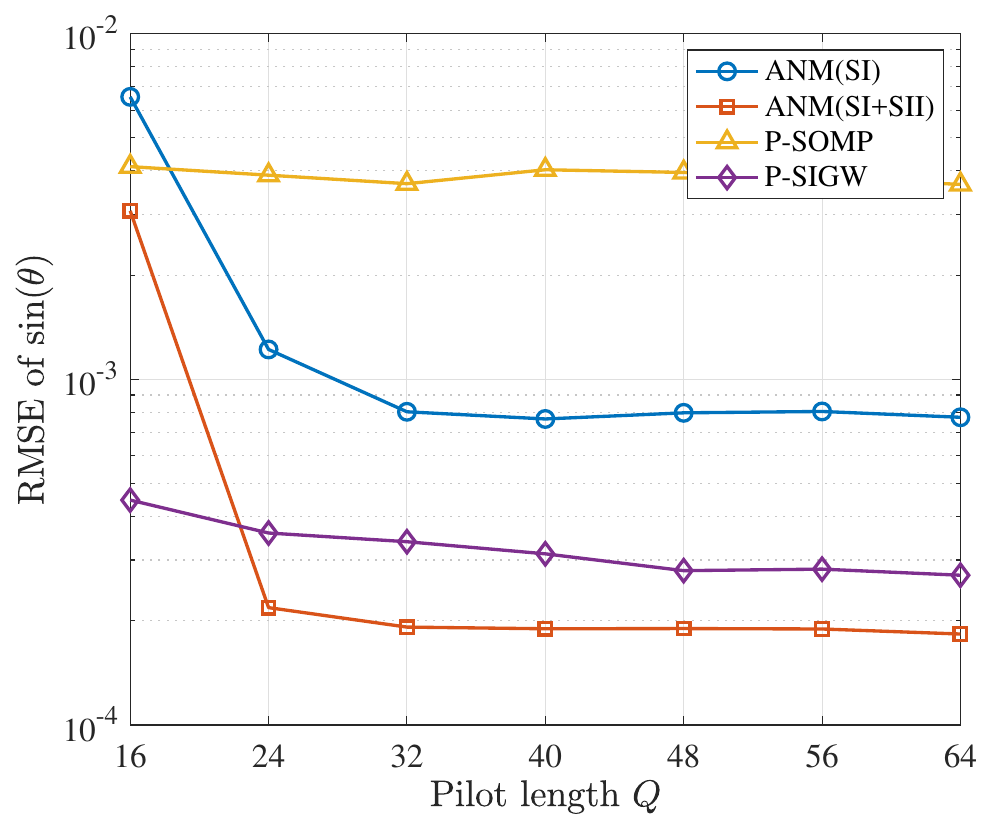}
        \caption{RMSE of $\sin(\theta)$.}  
        \end{subfigure}
        \begin{subfigure}[b]{0.6\columnwidth}
            \includegraphics[width=1\linewidth]{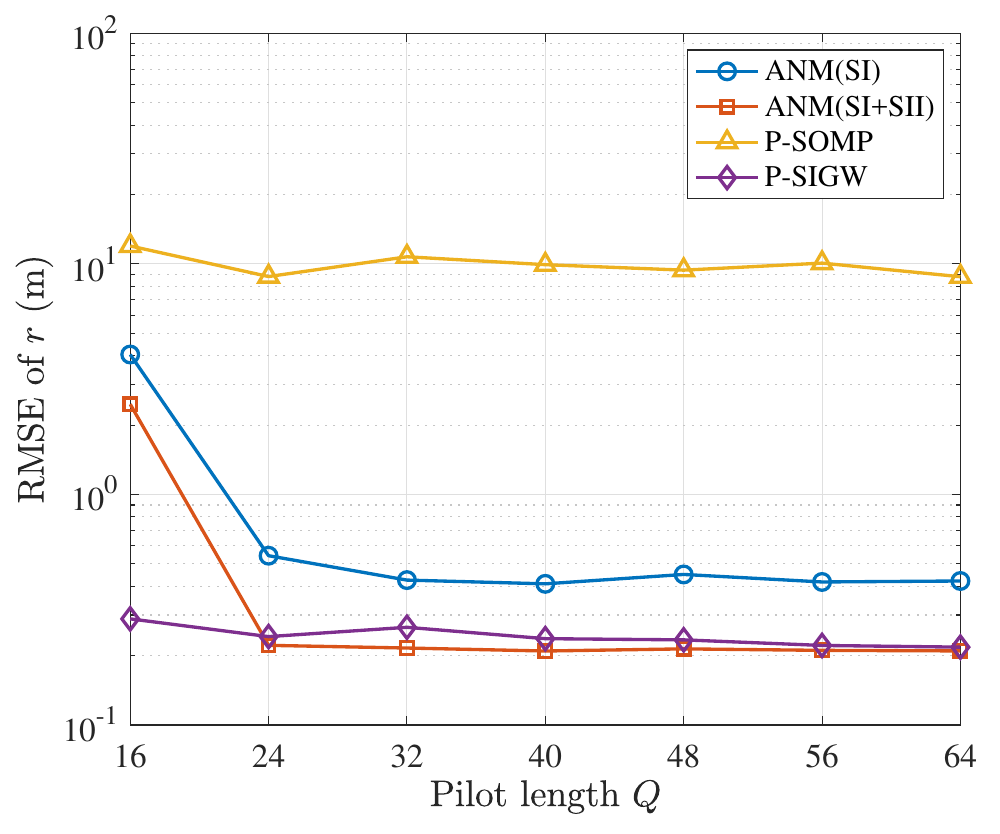}
        \caption{RMSE of $r$.}  
        \end{subfigure}
        \caption{NMSE performance of the proposed approaches and competing baselines with respect to different pilot length $Q$ when SNR = $30$ dB.}
        \label{fig:nmse_vs_Q}
\end{figure*}   

\begin{figure*}[t]
        \centering
        \begin{subfigure}[b]{0.6\columnwidth}
        \includegraphics[width=1\linewidth]{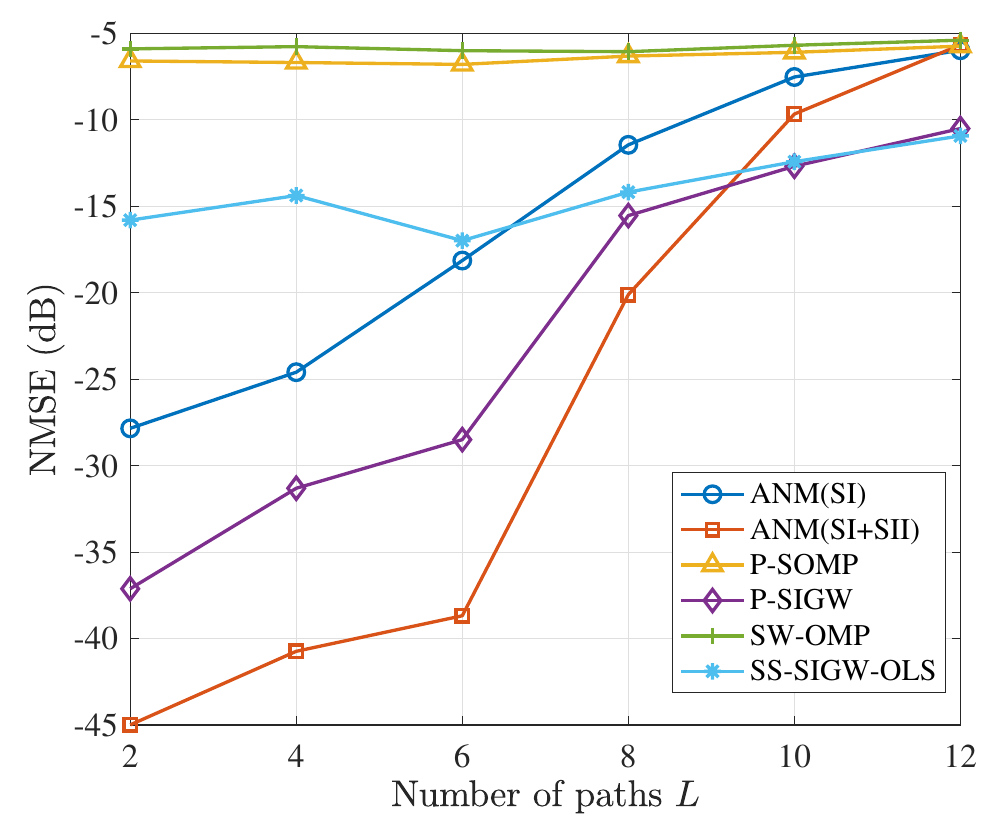}
        \caption{NMSE of $\mathbf{h}$ with $r\in[60,80]\text{m}$.}   
        \end{subfigure}
        \begin{subfigure}[b]{0.6\columnwidth}
            \includegraphics[width=1\linewidth]{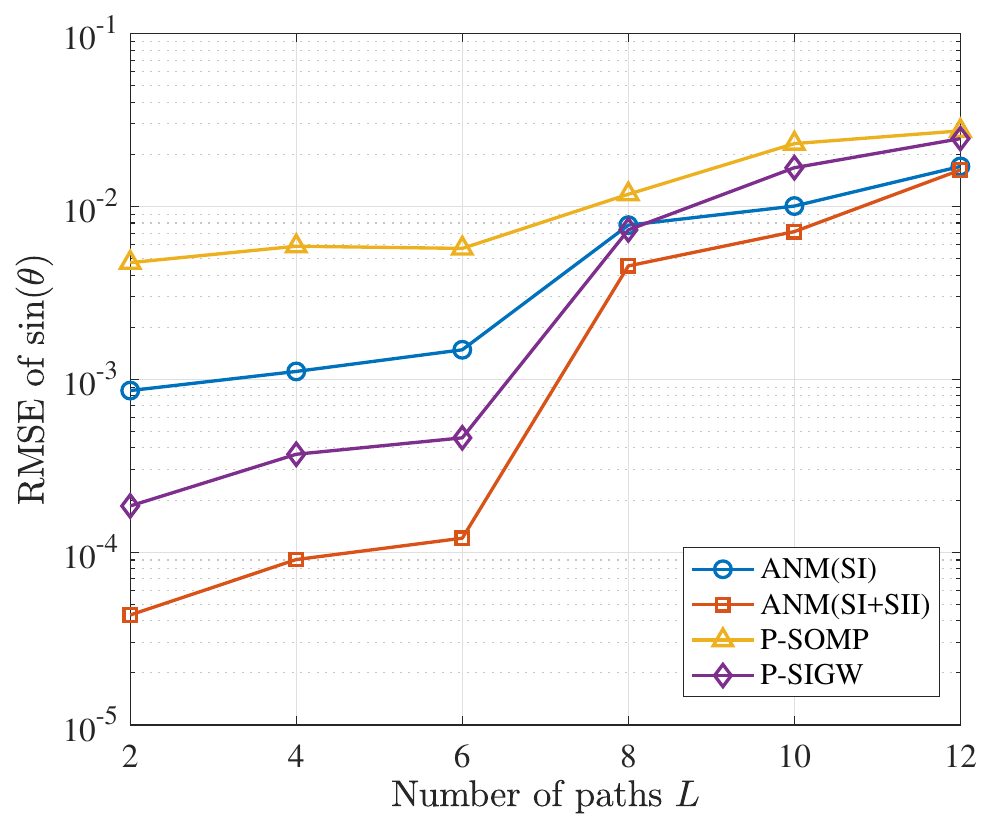}
        \caption{RMSE of $\sin(\theta)$ with $r\in[60,80]\text{m}$.}  
        \end{subfigure}
        \begin{subfigure}[b]{0.6\columnwidth}
            \includegraphics[width=1\linewidth]{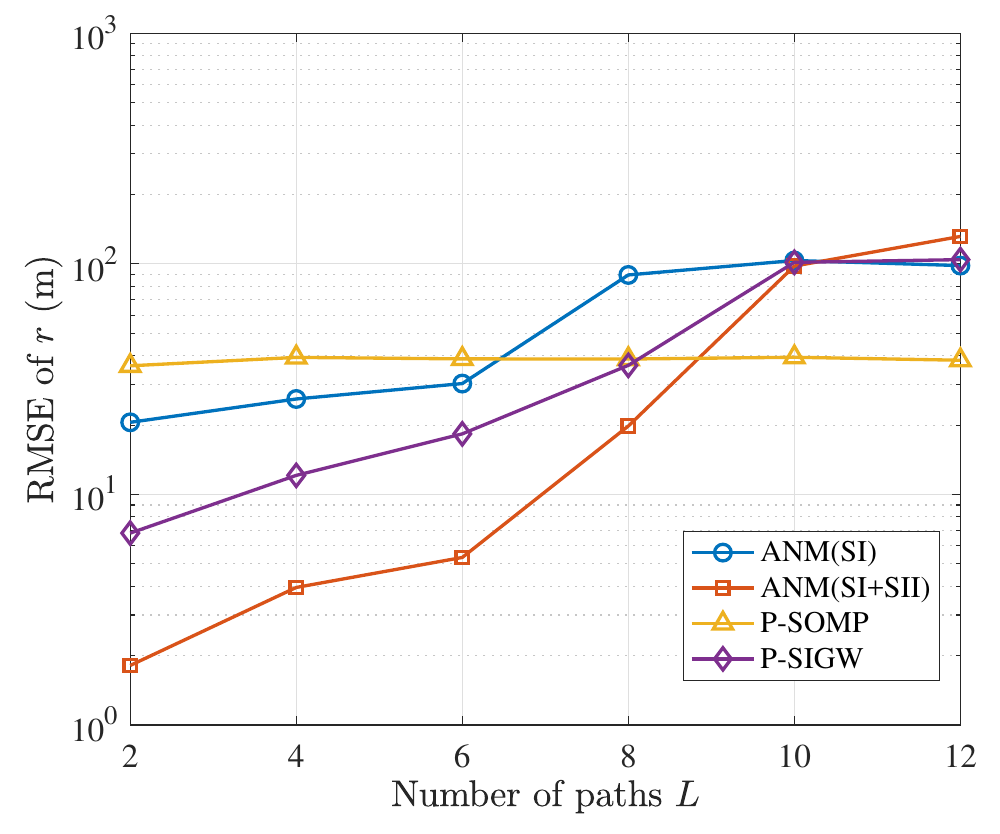}
        \caption{RMSE of $r$ with $r\in[60,80]\text{m}$.}  
        \end{subfigure}
        \caption{NMSE performance of the proposed approaches and competing baselines with respect to different number of propagation paths $L$ when SNR = $30$ dB.}
        \label{fig:nmse_vs_L}
\end{figure*} 

\subsection{Performance under Different SNR Levels}
\label{subsec:nmse_vs_snr}

We now evaluate channel, angle and range estimation performance across various SNR levels and with different range configurations. As shown in
Figs. \ref{fig:nmse_vs_SNR}(a) and \ref{fig:nmse_vs_SNR}(d), the NMSE of ANM(SI) decreases monotonically with increasing SNR, whereas the on-grid SW-OMP and P-SOMP algorithms exhibit error floors at moderate-to-high SNR due to basis mismatch. For SNR $>$ 5 dB, ANM(SI) significantly outperforms the on-grid methods, demonstrating the advantage of the proposed gridless super-resolution approach. By incorporating the refinement step in Stage-II of Algorithm~\ref{alg}, ANM(SI+SII) achieves lower NMSE than ANM(SI) and substantially outperforms the off-grid SS-SIGW-OLS and P-SIGW algorithms when SNR $>$ 15 dB.
In the low SNR regime ($-5$ to $5$ dB), ANM(SI) performs slightly worse than the polar-domain P-SOMP, and ANM(SI+SII) is marginally inferior to P-SIGW when SNR $<15$ dB. As indicated by Figs. \ref{fig:nmse_vs_SNR}(b) and \ref{fig:nmse_vs_SNR}(e), this behavior likely stems from less accurate Stage-I angle estimates at low SNR, which limits the effectiveness of Stage-II refinement.

Figs. \ref{fig:nmse_vs_SNR}(b) and \ref{fig:nmse_vs_SNR}(e)  illustrate the RMSE of angle estimation versus SNR. Both ANM(SI) and ANM(SI+SII) exhibit decreasing RMSE with increasing SNR. Notably, ANM(SI+SII) achieves substantial improvement at moderate and high SNR, attaining the lowest RMSE among all methods when SNR $\ge$ 15 dB.
In contrast, the on-grid P-SOMP suffers from an SNR-insensitive error floor due to basis mismatch. Although the off-grid P-SIGW improves with SNR, it eventually saturates at a higher RMSE than ANM(SI+SII). 
 Figs. \ref{fig:nmse_vs_SNR}(c) and \ref{fig:nmse_vs_SNR}(f) present the range estimation RMSE, which follows a similar trend to angle estimation. Across all tasks, ANM(SI+SII) provides significant additional gains over ANM(SI). These results confirm that Stage-II effectively mitigates residual angle and range biases arising from the second-order approximation of the near-field steering vector, leading to more accurate near-field parameter estimation.

\subsection{Performance with Different Range Parameters}
\label{subsec:nmse_vs_distance}

We now investigate the impact of the distance between the scatterers and the BS on the performance of the proposed near-field channel estimation method. The distance varies from 5m to 120m, covering the near-field region and extending toward the far-field boundary. In the simulations, we set SNR = $30$ dB. 
Fig.~\ref{fig:nmse_vs_Distance}(a) shows the NMSE of channel estimation versus distance. The NMSE of ANM(SI) remains nearly constant at approximately $-25$ dB across the entire range, indicating that the Stage-I reconstruction is largely insensitive to user location. In contrast, ANM(SI+SII) achieves substantially lower NMSE once the distance exceeds 20m, demonstrating the effectiveness of the Stage-II refinement when the user is not extremely close to the BS. The on-grid P-SOMP suffers from a persistent high error floor due to basis mismatch, whereas SW-OMP and SS-SIGW-OLS improve with increasing distance but remain ineffective in the deep near-field region.

The angle estimation results in Fig.~\ref{fig:nmse_vs_Distance}(b) follow a similar trend. ANM(SI+SII) consistently achieves the lowest RMSE across all distances, while P-SOMP exhibits the largest errors. The off-grid P-SIGW method outperforms the on-grid algorithms but still falls behind ANM(SI+SII). The range estimation performance in Fig.~\ref{fig:nmse_vs_Distance}(c) exhibits the expected distance-dependent behavior of near-field localization. As the distance increases, the RMSEs of ANM(SI), ANM(SI+SII), and P-SIGW grow approximately linearly, reflecting the increasing difficulty of accurate range recovery as the spherical wavefront gradually transitions toward planar propagation. Nevertheless, ANM(SI+SII) consistently outperforms ANM(SI) and significantly surpasses both P-SOMP and P-SIGW over the entire distance range.



\subsection{Performance with Different Pilot Lengths}
\label{subsec:nmse_vs_Q}

We next evaluate the impact of pilot length $Q$ on the performance of the proposed near-field channel estimation method. The pilot length varies from
 $Q=16$ to $Q=64$, corresponding to different training overheads. We fix the SNR at 30 dB. As shown in
 Fig.~\ref{fig:nmse_vs_Q}(a), the NMSEs of both ANM(SI) and ANM(SI+SII) improve steadily as $Q$ increases, and the curves gradually saturate for larger $Q$. These results demonstrate that the proposed methods effectively exploit additional pilot observations. In contrast, the on-grid P-SOMP 
 exhibits only marginal improvement with increasing pilot lengths, whereas SW-OMP and SS-SIGW-OLS maintain nearly constant and significantly higher NMSE across all values of $Q$.
The angle and range estimation results in Figs.~\ref{fig:nmse_vs_Q}(b) and \ref{fig:nmse_vs_Q}(c) follow a similar trend. Both ANM(SI) and ANM(SI+SII) benefit substantially from longer pilot sequences, further confirming that the proposed algorithms capture the spherical wavefront structure more accurately as more observations become available. 

\subsection{Performance with Different Number of Propagation Paths}
\label{subsec:nmse_vs_L}
We finally investigate the robustness of the proposed algorithms to multipath scenarios by varying the number of propagation paths $L$ from 2 to 12. As shown in
 Fig.~\ref{fig:nmse_vs_L}(a), the channel estimation NMSE of both ANM(SI) and ANM(SI+SII) increases with $L$, reflecting the greater difficulty of resolving a larger number of closely spaced paths. ANM(SI) maintains a clear advantage over the on-grid algorithms (SW-OMP and P-SOMP) for all tested values of $L$, while ANM(SI+SII) consistently achieves the lowest NMSE, outperforming the off-grid P-SIGW and SS-SIGW-OLS methods when $L\le 9$.

Fig.~\ref{fig:nmse_vs_L}(b) shows that the RMSEs of angle estimation degrade as $L$ increases for all algorithms. Nevertheless, ANM(SI+SII) delivers the highest accuracy across the entire range of $L$. When $L\ge 8$, the angle estimation performance of ANM(SI) and ANM(SI+SII) becomes comparable, and both significantly outperform P-SOMP and P-SIGW.
Finally, Fig.~\ref{fig:nmse_vs_L}(c) presents the RMSE of range estimation. We see that ANM(SI+SII) provides superior range accuracy for small-to-moderate values of $L$. As $L$ increases further, however, the benefit of the Stage-II refinement diminishes, and the performance of ANM(SI+SII) gradually approaches that of ANM(SI) and P-SIGW. Interestingly, P-SOMP exhibits an almost constant range RMSE across all $L$. 
This is because P-SOMP produces poor range estimates, which are essentially the maximum range parameter in the polar-domain dictionary.

\section{Conclusion}
\label{sec:conclusion}

This paper presented a super-resolution approach for near-field channel estimation in XL-MIMO systems. We showed that the near-field channel can be expressed as a set of complex exponentials modulated by unknown waveforms and demonstrated that these modulating waveforms lie in a common low-dimensional DCR subspace. By exploiting both the channel sparsity and subspace model, we formulated the channel estimation problem as a regularized atomic norm minimization, which can be solved via semidefinite programming. Theoretical recovery guarantees are established, showing that the proposed method can reliably reconstruct the underlying structured low-rank matrix, and hence the near-field channel, in the presence of noise. Extensive simulations confirmed the significant performance gains of the proposed approach over existing angular-domain and polar-domain methods in both channel estimation and joint angle-range parameter recovery. In the future, we plan to develop computationally efficient algorithms for the proposed approach and validate this approach using real-world measurement data.


\begin{appendix}
\numberwithin{proposition}{subsection} 
\numberwithin{lem}{subsection} 
\numberwithin{corollary}{subsection} 
\numberwithin{remark}{subsection} 
\numberwithin{equation}{subsection}	

\vspace{-0.2cm}
\subsection{Proof of Proposition \ref{Prop1}}
\label{app:Proof00}
Fix $\phi\in[0,\phi_{\max}]$ and let $\widetilde{\phi}_k$ be its nearest grid point such that $|\phi-\widetilde{\phi}_k|\leq \frac{\delta}{2}$. Define the normalized correlation function
    \begin{equation}
        \rho_N(\Delta \phi) = \langle\mathbf{u}(\phi), \mathbf{u}(\phi + \Delta \phi) \rangle = \frac{1}{N} \sum_{n=0}^{N-1} e^{j2\pi \Delta \phi n^2}.
    \end{equation}
    Using $|e^{j\theta} -1|\leq |\theta|$, we have
    \begin{equation}
        \begin{split}
            |\rho_N(\Delta \phi) -1| &= \lvert\frac{1}{N}\sum_{n=0}^{N-1} (e^{j2\pi\Delta\phi n^2} -1) \rvert \\ & \leq \frac{1}{N} \sum_{n=0}^{N-1} 2\pi |\Delta\phi|n^2 
            = 2\pi |\Delta\phi| \frac{1}{N}\sum_{n=0}^{N-1} n^2 \\
             & = 2\pi |\Delta\phi|\frac{(N-1)(2N-1)}{6},
        \end{split}
    \end{equation}
    where the last equality uses the identity $\sum_{n=0}^{N-1} n^2 = N(N-1)(2N-1)/6$.
    Therefore, the real part of $\rho_N(\Delta \phi)$ satisfies
    \begin{equation}
        \text{Re}(\rho_N(\Delta \phi)) \geq 1 - 2\pi |\Delta\phi|\frac{(N-1)(2N-1)}{6}.
    \end{equation}
    Thus, we obtain
    \begin{equation}
    \begin{split}
        \|\mathbf{u}(\phi) - \mathbf{u}(\phi + \Delta\phi)\|_2 &= \sqrt{2 - 2\cdot\text{Re}(\rho_N(\Delta \phi))} \\
        &\leq \sqrt{4\pi |\Delta\phi|\frac{(N-1)(2N-1)}{6}}.
    \end{split}
    \end{equation}
    Defining $\Delta\phi =  \widetilde{\phi}_k - \phi$, i.e., $|\Delta\phi|\leq \frac{\delta}{2}$, we have
    \begin{equation}\label{eqn: dist_u}
         \|\mathbf{u}(\phi) - \mathbf{u}(\widetilde{\phi}_k)\|_2 \leq \sqrt{\frac{\pi \delta(N-1)(2N-1)}{3}}.
    \end{equation}
    Since $\mathbf{u}(\widetilde{\phi}_k)\in\mathcal V$, the best approximation error is no larger than the error bound in (\ref{eqn: dist_u}), which proves \eqref{eqn:dist_bound_exact}.
    To ensure the right-hand side of \eqref{eqn:dist_bound_exact} is at most $\epsilon$, the lower bound of $P$ is given by \eqref{eqn:P_bound_exact} when $\delta=\phi_{\max}/(P-1)$.
    Finally, inserting $\phi_{\max}$ from Lemma \ref{lemma1} into \eqref{eqn:P_bound_exact} gives \eqref{eqn:P_bound_phi}.

\vspace{-0.2cm}
\subsection{Proof of Theorem \ref{thm1}}
\label{app:Proof1}
Recall that if a minimizer $\widehat{\mathbf{X}}$ exists, it should satisfy the subgradient condition
    \begin{equation}
        0\in \partial \|\widehat{\mathbf{X}}\|_\mathcal{A} + \frac{1}{\tau} \mathcal{P}^\ast_y(\mathcal{P}_y(\widehat{\mathbf{X}})-\mathbf{y}).
    \end{equation}
    There exists a dual variable $\mathbf{S}\in \partial \|\widehat{\mathbf{X}}\|_\mathcal{A}$ with $\|\mathbf{S}\|^\ast_\mathcal{A} \leq 1$ and $\langle \mathbf{S}, \widehat{\mathbf{X}} \rangle_\mathbb{R} = \|\widehat{\mathbf{X}}\|_\mathcal{A}$.
    Thus, we have $\mathbf{S} = -\frac{1}{\tau} \mathcal{P}^\ast_y(\mathcal{P}_y(\widehat{\mathbf{X}})-\mathbf{y}) $.

    For simplicity, denote $\Delta =\widehat{\mathbf{X}} - \mathbf{X}_h$ and $\mathbf{r}=\mathcal{P}_y(\widehat{\mathbf{X}})-\mathbf{y}=\mathcal{P}_y(\Delta)-\mathbf{n}$. 
    By convexity of the atomic norm, we have
    \begin{equation}
        \|\mathbf{X}_h\|_\mathcal{A} \geq \|\widehat{\mathbf{X}}\|_\mathcal{A} + \langle \mathbf{S}, \mathbf{X}_h - \widehat{\mathbf{X}}\rangle = \|\widehat{\mathbf{X}}\|_\mathcal{A} - \frac{1}{\tau} \langle \mathcal{P}_y^\ast(\mathbf{r}), \mathbf{X}_h - \widehat{\mathbf{X}} \rangle_\mathbb R
    \end{equation}
    Using the identity of the adjoint operation $\langle \mathbf{z}, \mathcal{P}_y(\mathbf{X})\rangle_{\mathbb{R}}=\langle \mathcal{P}_y^{\ast}(\mathbf{z}), \mathbf{X}\rangle_{\mathbb{R}}$, we have
    \begin{equation}
    \begin{split}
         \|\mathbf{X}_h\|_\mathcal{A} &\geq  \|\widehat{\mathbf{X}}\|_\mathcal{A} - \frac{1}{\tau}\langle \mathbf{r}, \mathcal{P}_y(\mathbf{X}_h - \widehat{\mathbf{X}}) \rangle_\mathbb R\\
         & = \|\widehat{\mathbf{X}}\|_\mathcal{A} + \frac{1}{\tau} \|\mathcal{P}_y(\Delta)\|_2^2 - \frac{1}{\tau} \langle \mathbf{n}, \mathcal{P}_y(\Delta) \rangle_\mathbb R, 
    \end{split}
    \end{equation}
    which is equivalent to 
    \begin{equation}\label{eq:proof_results}
        \frac{1}{2\tau} \|\mathcal{P}_y(\Delta)\|_2^2 \leq \frac{1}{2}\left(\|\mathbf{X}_h\|_\mathcal{A} -  \|\widehat{\mathbf{X}}\|_\mathcal{A} \right) + \frac{1}{2\tau} \langle \mathbf{n}, \mathcal{P}_y(\Delta) \rangle_\mathbb R.
    \end{equation}
    By applying Hölder’s inequality for dual norms, we have
    \begin{equation}\label{eq:noise_inequal}
        \begin{split}
            |\langle \mathbf{n}, \mathcal{P}_y(\Delta) \rangle_\mathbb R| &= |\langle\mathcal{P}_y^\ast(\mathbf{n}), \Delta \rangle_\mathbb R| \leq \|\mathcal{P}_y^\ast(\mathbf{n})\|_\mathcal{A}^\ast \|\Delta\|_\mathcal{A} \\
            &\leq \|\mathcal{P}_y^\ast(\mathbf{n})\|_\mathcal{A}^\ast \left(\|\widehat{\mathbf{X}}\|_\mathcal{A} + \|\mathbf{X}_h\|_\mathcal{A} \right).
        \end{split}
    \end{equation}
    If the condition $\| \mathcal{P}_y^{\ast}(\mathbf{n})\|_\mathcal{A}^{\ast} \le\ \tau$ holds, inserting (\ref{eq:noise_inequal}) into (\ref{eq:proof_results}) leads to (\ref{eq:measured-bound}) in Theorem \ref{thm1}. 
    
    \qed

\color{black}
\subsection{Proof of Theorem \ref{newthm}}
\label{app:newthm}
We bound $\mathbb{E} \| \mathcal{P}_y^{\ast}(\mathbf(n)\|_\mathcal{A}^{\ast}$ by following a similar procedure in Appendix C and Appendix D of \cite{Bhaskar-TSP13}. Notice that 
\begin{equation*}
\begin{aligned}
(\| \mathcal{P}_y^{\ast}(\mathbf{n})\|_\mathcal{A}^{\ast})^2 & = \sup_{\omega\in[0,1)} \|\mathcal{P}_y^{\ast}(\mathbf{n})\overline{\mathbf{a}(\omega)}\|_2^2 \\
& = \sup_{\omega\in[0,1)} \sum_{p=0}^{P-1} \left|\sum_{n=0}^{N-1} [\mathbf{A}^H \mathbf{n}]_n [\mathbf{q}_n]_p e^{-j2\pi \omega n} \right|^2 \\
& = \sup_{\omega\in[0,1)} \sum_{p=0}^{P-1} \sum_{n=0}^{N-1} \sum_{m = 0}^{N-1} [\mathbf{A}^H \mathbf{n}]_n \overline{[\mathbf{A}^H \mathbf{n}]_m} \\ &~~~~~~~~~~~\cdot [\mathbf{q}_n]_p \overline{[\mathbf{q}_m]_p} e^{j2\pi\omega(m-n)}. 
\end{aligned}
\end{equation*}
Define 
\begin{equation}
\label{eq:d2}
\begin{aligned}
W_N(e^{j2\pi \omega}) & = \sum_{p=0}^{P-1} \sum_{n=0}^{N-1} \sum_{m = 0}^{N-1} [\mathbf{A}^H \mathbf{n}]_n \overline{[\mathbf{A}^H \mathbf{n}]_m}  \\ 
&~~~~~~~\cdot [\mathbf{q}_n]_p \overline{[\mathbf{q}_m]_p} e^{j2\pi\omega(m-n)}.
\end{aligned}
\end{equation}
For any $\omega_1, \omega_2 \in [0, 1)$, we have 
\begin{equation}
\label{eq:bernstein}
\begin{aligned}
& W_N(e^{j2\pi \omega_1}) - W_N(e^{j2\pi \omega_2}) \\
\leq & |e^{j2\pi \omega_1} - e^{j2\pi\omega_2}| \sup_{\omega\in[0,1)} \left|W_N'(e^{j2\pi \omega})\right| \\
= & 2 | \sin(\pi (\omega_1 - \omega_2))|\sup_{\omega\in[0,1)} \left|W_N'(e^{j2\pi \omega})\right| \\
\leq & 2\pi |\omega_1 - \omega_2| \sup_{\omega\in[0,1)} \left|W_N'(e^{j2\pi \omega})\right| \\
\leq & 4\pi N |\omega_1 - \omega_2|   \sup_{\omega\in[0,1)}  W_N(e^{j2\pi \omega}),
\end{aligned}
\end{equation}
where the first inequality uses the mean value theorem and the last inequality follows from Bernstein's theorem (see, for example, Theorem 3 in Appendix C of \cite{Bhaskar-TSP13}) and the fact that $\underset{\omega\in[0,1)}{\sup} W_N(e^{j2\pi \omega})$ from (\ref{eq:d2}) is nonnegative. Defining the grid values $0, \frac{1}{T}, \cdots, \frac{T-1}{T}$ and using (\ref{eq:bernstein}), it yields 
\begin{equation}
\begin{aligned}
& \sup_{\omega\in[0,1)} W_N(e^{j2\pi \omega}) \leq \max_{t=0,\cdots,T-1} W_N(e^{j2\pi \frac{t}{T}}) \\
& ~~~~~~~~~~~~~~~~~~~~~~~+ \frac{2\pi N}{T}\sup_{\omega\in[0,1)} W_N(e^{j2\pi \omega}), 
\end{aligned}
\end{equation}
which leads to 
\begin{equation}
\label{eq:D4}
\begin{aligned}
\sup_{\omega\in[0,1)} W_N(e^{j2\pi \omega}) & \leq \left(1 - \frac{2\pi N}{T}\right)^{-1} \max_{t=0,\cdots,T-1} W_N(e^{j2\pi \frac{t}{T}}) \\
& \leq \left(1 + \frac{4\pi N}{T}\right) \max_{t=0,\cdots,T-1} W_N(e^{j2\pi \frac{t}{T}}),
\end{aligned}
\end{equation}
where the last inequality holds when $T \geq 4\pi N$. Since $(\| \mathcal{P}_y^{\ast}(\mathbf{n})\|_\mathcal{A}^{\ast})^2 = \underset{\omega\in[0,1)}{\sup} W_N(e^{j2\pi \omega})$, we bound $\mathbb{E}(\| \mathcal{P}_y^{\ast}(\mathbf{n})\|_\mathcal{A}^{\ast})^2$ by upper bounding $\mathbb{E} \underset{t=0,\cdots,T-1}{\max} W_N(e^{j2\pi \frac{t}{T}})$, where the expectation is with respect to the noise $\mathbf{n}$. Note that 
\begin{equation}
\label{eq:D5}
\begin{aligned}
& \mathbb{E} \max_{t=0,\cdots,T-1} W_N(e^{j2\pi \frac{t}{T}}) \\
= & \mathbb{E} \max_{t=0,\cdots,T-1} \sum_{p=0}^{P-1} \left|\sum_{n=0}^{N-1} [\mathbf{A}^H \mathbf{n}]_n [\mathbf{q}_n]_p e^{-j2\pi \frac{t}{T} n} \right|^2 \\
\leq & \sum_{p=0}^{P-1} \mathbb{E} \max_{t=0,\cdots,T-1} \left|\sum_{n=0}^{N-1} [\mathbf{A}^H \mathbf{n}]_n [\mathbf{q}_n]_p e^{-j2\pi \frac{t}{T} n} \right|^2.
\end{aligned}
\end{equation}
Define $x_{p, t} = \sum_{n=0}^{N-1} [\mathbf{A}^H \mathbf{n}]_n [\mathbf{q}_n]_p e^{-j2\pi \frac{t}{T} n}$. It is easy to see $
\mathbb{E}(x_{p, t}) = 0
$.  
Denote $[\mathbf{\widetilde{q}}_n]_{p, t} = [\mathbf{q}_n]_{p} e^{-j2\pi \frac{t}{T} n}$. For notational convenience, we also define $\mathbf{a'}_m^T$ as the $m$th row of $\mathbf{A}$ and $\mathbf{g'}_{p, t}$ a length-$N$ column vector with its $n$th entry being $[\mathbf{\widetilde{q}}_n]_{p, t}$. Then 
\begin{equation}
\begin{aligned}
\text{Var}(x_{p, t}) & = \text{Var}\left(\sum_{m=0}^{M-1} \sum_{n=0}^{N-1} \overline{[\mathbf{A}]}_{m, n} [\mathbf{\widetilde{q}}_n]_p [\mathbf{n}]_m\right) \\
& = \text{Var}\left(\sum_{m=0}^{M-1} 
\mathbf{a'}^H_m \mathbf{g'}_{p, t} [\mathbf{n}]_m\right) \\
& = \sum_{m=0}^{M-1} |\mathbf{a'}^H_m \mathbf{g'}_{p, t}|^2\sigma^2 = \| \overline{\mathbf{A}}\mathbf{g'}_{p,t}\|_2^2 \sigma^2
\end{aligned}
\end{equation}
Denote $\sigma_{p,t}^2 = \| \overline{\mathbf{A}} \mathbf{g'}_{p,t}\|_2^2 \sigma^2$ and $\sigma_p^2 = \|\mathbf{A}\|_2^2 \|\mathbf{g}_p\|_2^2 \sigma^2$, where $\mathbf{g}_p$ is the $p$th column of $\mathbf{G}$. Thus, $x_{p, t}\sim \mathcal{CN}(0, \sigma_{p,t}^2)$ and $\frac{2|x_{p, t}|^2}{\sigma_{p,t}^2}$ is a chi-squared random variable with two degrees of freedom. Furthermore, we have 
\begin{equation}
\label{eq:D7}
\begin{aligned}
& \mathbb{E} \left[ \max_{0, \cdots, T-1} 2 |x_{p, t}|^2\right] = \int_0^{\infty} \mathbb{P} \left\{ \max_{t=0, \cdots, T-1} 2|x_{p,t}|^2 \geq t \right\} dt \\
= & \int_0^{\delta_p}  \mathbb{P} \left\{ \max_{t=0, \cdots, T-1} 2|x_{p,t}|^2 \geq t \right\} dt \\
& ~~~~~~~~~~~~~~~~~+\int_{\delta_p}^{\infty} \mathbb{P} \left\{ \max_{t=0, \cdots, T-1} 2|x_{p,t}|^2 \geq t \right\} dt \\
\leq & \delta_{p} + T \int_{\delta_{p}}^{\infty} \mathbb{P}\left\{  \frac{2|x_{p,t}|^2}{\sigma_{p,t}^2} \geq \frac{t}{\sigma_{p,t}^2} \right\} dt \\
\leq & \delta_{p} + T \int_{\delta_{p}}^{\infty} \mathbb{P}\left\{  \frac{2|x_{p,t}|^2}{\sigma_{p,t}^2} \geq \frac{t}{\sigma_{p}^2} \right\} dt \\
= & \delta_p + T \int_{\delta_p}^{\infty} e^{-\frac{t}{2\sigma_p^2}} dt = \delta_p + 2T \sigma_p^2 e^{-\frac{\delta_p}{2\sigma_p^2}},
\end{aligned}
\end{equation}
where the second inequality above uses the fact that $\sigma_{p,t}^2 \leq \sigma_{p}^2$. 
By choosing $\delta_p = 2\sigma_p^2 \log T$ and $T = 4\pi N \log N$, we have 
\begin{equation*}
\begin{aligned}
& \mathbb{E}(\| \mathcal{P}_y^{\ast}(\mathbf{n})\|_\mathcal{A}^{\ast})  \\
\leq & \sqrt{ 1 + \frac{1}{\log N}} \sqrt{\left(\frac{1}{2} \sum_{p=0}^{P-1} \mathbb{E} \left[ \max_{0, \cdots, T-1} 2 |x_{p, t}|^2\right] \right)} \\
= &  \sqrt{ 1 + \frac{1}{\log N}} \sqrt{\sum_{p=0}^{P-1} \left(\sigma_p^2 \log (4\pi N \log N) + \sigma_p^2 \right)} \\
= &  \sqrt{ 1 + \frac{1}{\log N}} \sqrt{\log N + \log (4\pi \log N) + 1}\sqrt{\sum_{p=0}^{P-1}\sigma_p^2 } \\
= & \sqrt{ 1 + \frac{1}{\log N}} \sqrt{\log N + \log (4\pi \log N) + 1} \|\mathbf{A}\|_2 \|\mathbf{G}\|_F \sigma
\end{aligned}
\end{equation*}
where the first inequality above follows directly from (\ref{eq:D4}), (\ref{eq:D5}) and (\ref{eq:D7}). This completes the proof. 
\hfill$\square$

\end{appendix}
 
\bibliographystyle{IEEEtran}
\bibliography{IEEEabrv,NearField}

\end{document}